\documentclass[journal]{IEEEtran}

\usepackage{scalerel}
\usepackage{times}
\usepackage{relsize}

\usepackage{enumitem}
\usepackage{float}
\usepackage{mathtools}
\usepackage{graphicx}
\usepackage{graphics}
\usepackage[dvips]{epsfig}
\usepackage{subfig}
\usepackage{verbatim}
\usepackage{graphicx,float,latexsym,amssymb,amsfonts,amsmath,amstext,times,epsfig,setspace}
\usepackage{algorithm}
\usepackage[utf8]{inputenc}
\usepackage[noend]{algpseudocode}
\usepackage{fancyhdr}
\usepackage{stfloats}
\usepackage{tabu}
\usepackage{booktabs}
\usepackage{amsmath}
\usepackage[noend]{algpseudocode}
\renewcommand{\algorithmiccomment}[1]{\bgroup\hfill\scriptsize//~#1\egroup}

\DeclareMathOperator*{\argmax}{arg\,max}
\DeclareMathOperator*{\argmin}{arg\,min}

\newcommand{\Exp}[1]{\mathbf{E}\left[{#1}\right]}

\newcommand{\Prob}[1]{\mathbf{Pr}\left[{#1}\right]}
\newcommand{\Probi}[2]{\mathbf{Pr}_{#1}\left[{#2}\right]}

\newcommand{\entropy}[1]{\mathbf{H}\left({#1}\right)}

\newcommand{\betap}{{\beta'}}

\newcommand{\liqsystem}{liquid system}
\newcommand{\Liqsystem}{Liquid system}
\newcommand{\liqrepairer}{liquid repairer}
\newcommand{\Liqrepairer}{Liquid repairer}
\newcommand{\advliqrepairer}{advanced liquid repairer}
\newcommand{\Advliqrepairer}{Advanced liquid repairer}

\newcommand{\tradsystem}{small code system}

\newcommand{\tradrepair}{reactive repair}

\newcommand{\Regframe}{Regenerating framework}
\newcommand{\Reglower}{Regenerating lower bound}
\newcommand{\Regrepairer}{Regenerating repairer}

\newcommand{\srcdata}{source data}
\newcommand{\Srcdata}{Source data}

\newcommand{\Afnc}{{\mathbf{A}}}
\newcommand{\bdata}{b}
\newcommand{\Bvar}{B}
\newcommand{\Cdata}{C}
\newcommand{\dvar}{d}
\newcommand{\dhvar}{\hat{d}}
\newcommand{\Ddata}{D}

\newcommand{\Fdata}{F}
\newcommand{\Fpdata}{F'}
\newcommand{\fvar}{f}
\newcommand{\fhvar}{\hat{f}}

\newcommand{\Gvar}{G}

\newcommand{\Gpvar}{\hat{G}}
\newcommand{\gsum}{{\rm gs}}

\newcommand{\Gseq}{{\rm Gseq}}
\newcommand{\Gpseq}{\hat{\rm G}{\rm seq}}
\newcommand{\hvar}{h}
\newcommand{\hhvar}{\hat{h}}
\newcommand{\kdata}{k}
\newcommand{\Kdata}{K}

\newcommand{\Lseq}{{\rm Lseq}}
\newcommand{\ndata}{n}
\newcommand{\Mdata}{M}
\newcommand{\Nsetp}{{\cal N}_{p}}

\newcommand{\Ndata}{N}
\newcommand{\Ppred}{P}
\newcommand{\pvar}{p}
\newcommand{\phvar}{\hat{p}}

\newcommand{\Qvar}{Q}

\newcommand{\Qseq}{{\rm Qseq}}
\newcommand{\Rfnc}{{\mathbf{R}}}
\newcommand{\Rpfnc}{{\mathbf{R}'}}
\newcommand{\rdata}{r}
\newcommand{\rpdata}{r'}
\newcommand{\Rdata}{R}
\newcommand{\Sdata}{S}
\newcommand{\thvar}{\hat{t}}
\newcommand{\thvarp}{\hat{t}^{\scaleto{+}{4pt}}}
\newcommand{\thvarm}{\hat{t}^{\scaleto{-}{4pt}}}
\newcommand{\tvar}{t}
\newcommand{\Tvar}{T}
\newcommand{\tvarp}{t^{\scaleto{+}{4pt}}}
\newcommand{\tvarm}{t^{\scaleto{-}{4pt}}}

\newcommand{\tseq}{{\rm tseq}}
\newcommand{\Tseq}{{\rm Tseq}}
\newcommand{\Thvar}{\hat{T}}

\newcommand{\Thvarm}{\hat{T}^{\scaleto{-}{4pt}}}

\newcommand{\Tpseq}{\hat{\rm T}{\rm seq}}

\newcommand{\Uvar}{U}

\newcommand{\Useq}{{\rm Useq}}
\newcommand{\Vdata}{V}
\newcommand{\xseq}{{\rm xseq}}

\newcommand{\xvar}{x}
\newcommand{\xhvar}{\hat{x}}
\newcommand{\Xvar}{X}
\newcommand{\Xset}{{\cal X}}
\newcommand{\yvar}{y}
\newcommand{\Yvar}{Y}
\newcommand{\yseq}{{\rm yseq}}
\newcommand{\Yseq}{{\rm Yseq}}
\newcommand{\Zvar}{Z}
\newcommand{\zvar}{z}

\newcommand{\epscore}{\epsilon_{c}}
\newcommand{\delcore}{\delta_{\scriptscriptstyle \rm c}}
\newcommand{\epsgeo}{\epsilon_{d}}
\newcommand{\delgeo}{\delta_{d}}

\newcommand{\deluni}{\delta_{u}}
\newcommand{\epspoi}{\epsilon}
\newcommand{\delpoi}{\delta}
\newcommand{\epsliq}{\epsilon}
\newcommand{\delliq}{\delta}

\newcommand{\erate}{{\cal E}}
\newcommand{\rrate}{{\cal R}}

\newcommand{\clen}{{\rm clen}}
\newcommand{\flen}{{\rm flen}}
\newcommand{\olen}{{\rm olen}}
\newcommand{\rlen}{{\rm rlen}}
\newcommand{\tlen}{{\rm tlen}}
\newcommand{\slen}{{\rm slen}}
\newcommand{\ulen}{{\rm ulen}}
\newcommand{\vlen}{{\rm vlen}}
\newcommand{\wlen}{{\rm wlen}}
\newcommand{\xlen}{{\rm xlen}}

\newcommand{\rcapacity}{capacity}

\newcommand{\erasurerate}{erasure rate}

\newcommand{\storeoverhead}{storage overhead}

\newcommand{\rrepairrate}{read rate}

\newcommand{\placegroup}{placement group}
\newcommand{\Placegroup}{Placement group}

\newcommand{\len}[1]{\lvert\lvert{#1}\rvert\rvert}
\newcommand{\abs}[1]{\left\lvert{#1}\right\rvert}
\newcommand{\ang}[1]{\langle{#1}\rangle}
\newcommand{\set}[1]{\left\{{#1}\right\}}

\newcommand{\bitval}{bit}

\newcommand{\lrepairer}{local-computation repairer}
\newcommand{\Lrepairer}{Local-computation repairer}

\newcommand{\inu}{\in_{\cal U}}

\newcommand{\nfail}{failure}
\newcommand{\Nfail}{Failure}
\newcommand{\nfseq}{failure sequence}

\newcommand{\helper}{helper}
\newcommand{\Helper}{Helper}
\newcommand{\primary}{primary}
\newcommand{\Primary}{Primary}

\newcommand{\timeseq}{timing sequence}

\newcommand{\idseq}{identifier sequence}

\newcommand{\identifier}{identifier}

\newcommand{\Ptdist}{Poisson failure distribution}

\newcommand{\lnifunction}{{\mathbf{lni}}}
\newcommand{\lndfunction}{{\mathbf{lnd}}}

\newcommand{\expp}[1]{10^{#1}}
\newcommand{\expm}[1]{10^{- #1}}

\newtheorem{theorem}{Theorem}[section]

\newtheorem{complemma}[theorem]{{\bf Compression Lemma}}
\newcommand{\Complemma}{Compression Lemma~\ref{compression lemma}}

\newtheorem{compcorollary}[theorem]{{\bf Compression Corollary}}
\newcommand{\Compcorollary}{Compression Corollary~\ref{compression corollary}}

\newtheorem{corelemma}[theorem]{{\bf Core Lemma}}
\newcommand{\Corelemma}{Core Lemma~\ref{core lemma}}

\newtheorem{coretheorem}[theorem]{{\bf Core Theorem}}
\newcommand{\Coretheorem}{Core Theorem~\ref{core theorem}}

\newtheorem{supertheorem}[theorem]{{\bf Supermartingale Theorem}}
\newcommand{\Supertheorem}{Supermartingale Theorem~\ref{super theorem}}

\newtheorem{geolemma}[theorem]{{\bf Distinct Failures Lemma}}
\newcommand{\Geolemma}{Distinct Failures Lemma~\ref{geo lemma}}

\newtheorem{unitheorem}[theorem]
{{\bf Uniform Failures Lower Bound Theorem}}
\newcommand{\Unitheorem}
{Uniform Failures Lower Bound Theorem~\ref{uniform theorem}}

\newtheorem{Poissontheorem}[theorem]
{{\bf Poisson Failures Lower Bound Theorem}}
\newcommand{\Poissonthm}
{Poisson Failures Lower Bound Theorem~\ref{Poisson theorem}}

\newtheorem{Liqperiodictheorem}[theorem]
{{\bf Liquid Periodic Failures Theorem}}
\newcommand{\Liqperiodicthm}
{Liquid Periodic Failures Theorem~\ref{Liqperiodic theorem}}

\newtheorem{Liqpoissontheorem}[theorem]
{{\bf Liquid Poisson Failures Theorem}}
\newcommand{\Liqpoissonthm}
{Liquid Poisson Failures Theorem~\ref{Liqpoisson theorem}}

\newtheorem{ALiqperiodictheorem}[theorem]
{{\bf Advanced Liquid Periodic Failures Theorem}}
\newcommand{\ALiqperiodicthm}
{Advanced Liquid Periodic Failures Theorem~\ref{ALiqperiodic theorem}}

\newtheorem{ALiqpoissontheorem}[theorem]
{{\bf Advanced Liquid Poisson Failures Theorem}}
\newcommand{\ALiqpoissonthm}
{Advanced Liquid Poisson Failures Theorem~\ref{ALiqpoisson theorem}}

\newenvironment{proof}[1][Proof]{\begin{trivlist}
\item[\hskip \labelsep {\bfseries #1}]}{\end{trivlist}}

\newcommand{\qed}{\nobreak \ifvmode \relax \else
      \ifdim\lastskip<1.5em \hskip-\lastskip
      \hskip1.5em plus0em minus0.5em \fi \nobreak
      \vrule height0.75em width0.5em depth0.25em\fi}

\begin{document}
\pagenumbering{gobble}

\title{Capacity bounds for distributed storage}
\thispagestyle{fancy}
\author{Michael~G.~Luby,~\IEEEmembership{~IEEE~Fellow,~ACM~Fellow} 
\IEEEcompsocitemizethanks{\IEEEcompsocthanksitem The author is with Qualcomm Technologies Inc., San Diego,
CA, 92121 USA e-mail: luby@qti.qualcomm.com, theluby@ieee.org.}
\thanks{Revised draft: April 11, 2018}
}

\fancyhead[LO]{\small Luby, Capacity bounds for distributed storage}
\fancyfoot[L]{\em Revised draft: April 11, 2018}
\renewcommand{\headrulewidth}{0pt}

\maketitle

\begin{abstract}

One of the primary objectives of a distributed storage system is to reliably store large
amounts of \srcdata\ for long durations using a large number $\Ndata$
of unreliable storage nodes, each with $\clen$ \bitval s of storage capacity.  
Storage nodes fail randomly over time and are replaced with nodes of equal capacity initialized to zeroes, and thus \bitval s are erased at some rate $\erate$. 
To maintain recoverability of the \srcdata,
a repairer continually reads data over a network from nodes at a rate $\rrate$, 
and generates and writes data to nodes based on the read data.

The {\em distributed storage \srcdata\ capacity} is the maximum
amount of \srcdata\ that can be reliably stored for long periods of time.
We prove the distributed storage \srcdata\ capacity asymptotically approaches
\begin{equation}
\label{fundamental eq}
\left(1-\frac{\erate}{2 \cdot \rrate}\right) \cdot \Ndata \cdot \clen
\end{equation}
as $\Ndata$ and $\rrate$ grow.

Equation~\eqref{fundamental eq} expresses a fundamental trade-off
between network traffic and storage overhead to reliably store \srcdata.
\end{abstract}

\begin{IEEEkeywords}
distributed information systems, data storage systems, data warehouses, information science, 
information theory, information entropy, error compensation, mutual information, channel capacity, channel coding, time-varying channels, error correction codes, Reed-Solomon codes, network coding, 
signal to noise ratio, throughput, 
distributed algorithms, algorithm design and analysis, 
reliability, reliability engineering, reliability theory, fault tolerance, redundancy, robustness,
failure analysis, equipment failure.
\end{IEEEkeywords}

\IEEEpeerreviewmaketitle

\section{Overview of practical systems}
\label{practical sec}

\IEEEPARstart{A}{ }distributed storage system generically consists of interconnected storage nodes,
where each node can store a large quantity of data.    A primary goal of a distributed
storage system is to reliably store as much \srcdata\ as possible for a long time. 

Commonly, distributed storage systems are built using relatively inexpensive
and generally not completely reliable hardware.
For example, nodes can go offline for periods of time (transient failure), in which case 
the data they store is temporarily unavailable, 
or permanently fail, in which case the data they store is permanently erased.   
Permanent \nfail s are not uncommon, and transient \nfail s are frequent.

Although it is often hard to accurately model \nfail s, 
an independent \nfail\ model can provide insight into 
the strengths and weaknesses of a practical system, 
and can provide a first order approximation to how a practical system operates.
In fact, one of the primary reasons practical storage systems are built using distributed infrastructure
is so that failures of the infrastructure are as independent as possible. 

Distributed storage systems generally allocate a 
fraction of their \rcapacity\ to \storeoverhead,
which is used to help maintain recoverability of \srcdata\ as \nfail s occur. 
Redundant data, which can be used to help recover lost \srcdata,
is generated from \srcdata\ and stored in addition to \srcdata.
A {\em repairer} reads stored data to
regenerate and restore lost data as \nfail s occur. 

For practical systems, \srcdata\ is generally  maintained 
at the granularity of {\em objects}, and erasure codes are 
used to generate redundant data for each object.  
For a $(n, k, r)$ erasure code, each object is segmented into $k$ source fragments, 
an encoder generates $r = n-k$ repair fragments from the $k$ source fragments, 
and each of these $n = k+r$ fragments is stored at a different node.  
An erasure code is MDS (maximum distance separable) 
if the object can be recovered from any $k$ of the $n$ fragments.

Replication is an example of a trivial MDS erasure code, i.e., 
each fragment is a copy of the original object.  
For example, triplication can be thought of as using the simple $(3,1,2)$ erasure code, 
wherein the object can be recovered from any one of the three copies.  
Many distributed storage systems use replication. 

Reed-Solomon codes \cite{CCauchy95}, \cite{CRizzo97}, \cite{RFC5510} are MDS codes
that are used in a variety of applications and are a popular choice for storage systems.
For example, \cite{Huang12} and~\cite{Ford10} use a $(9,6,3)$ Reed-Solomon code, 
and~\cite{Dimakis13} uses a $(14,10,4)$ Reed-Solomon code.  These are
examples of {\em \tradsystem s}, i.e., systems that use small values of $n$, $k$ and $r$.

Since a small number $r+1$ of \nfail s can cause \srcdata\ loss for \tradsystem s,
{\em \tradrepair} is used,  i.e., the repairer operates as quickly as practical to 
regenerate fragments lost from a node that permanently fails before another node fails,
and typically reads $k$ fragments to regenerate each lost fragment.
Thus, the peak \rrepairrate\ is higher than the average \rrepairrate, 
and the average \rrepairrate\ is $k$ times the \nfail\ \erasurerate. 

As highlighted in~\cite{Dimakis13}, the \rrepairrate\ needed 
to maintain \srcdata\ recoverability for \tradsystem s can be substantial.
Modifications of standard erasure codes have been  
designed for storage systems to reduce this rate,  
e.g., {\em local reconstruction codes} \cite{Gopalan12}, \cite{Dimakis13}, 
and {\em regenerating codes} \cite{Dimakis07}, \cite{Dimakis10}.  
Some versions of local reconstruction codes have been used in deployments, e.g., by Microsoft Azure.

There are additional issues that make the design of \tradsystem s complicated.
For example, {\em \placegroup s}, each mapping $n$ fragments to $n$ of the $\Ndata$ nodes,  
are used to determine where fragments for objects are stored. 
An equal amount of object data should be assigned to each \placegroup, 
and an equal number of \placegroup s should map a fragment to each node.
For \tradsystem s, Ceph~\cite{Ceph} recommends $\frac{100 \cdot \Ndata}{n}$ \placegroup s,
i.e., 100 \placegroup s map a fragment to each node. 
A \placegroup\ should avoid mapping fragments to nodes with
correlated failures, e.g., to the same rack.
Pairs of \placegroup s should avoid mapping fragments to the same pair of nodes.
\Placegroup s are continually remapped as nodes fail and are added. 
These and other issues make the design of \tradsystem s challenging. 

The paper~\cite{Luby16} introduces {\em \liqsystem s},
which use erasure codes with large values of $n$, $k$ and $r$.  
For example, $n=\Ndata$ and a fragment is assigned to each node for each object,
i.e., only one \placegroup\ is used for all objects.
The RaptorQ code \cite{CRaptorQ11},~\cite{RFC6330} is an example of 
an erasure code that is suitable for a \liqsystem, 
since objects with large numbers of fragments can be encoded and decoded
efficiently in linear time.
  
Typically $r$ is large for a \liqsystem, thus \srcdata\
is unrecoverable only when a large number of nodes fail.
A \liqrepairer\ is lazy, i.e., repair operates to 
slowly regenerate fragments erased from nodes that have permanently failed.
The repairer reads $k$ fragments for each object to regenerate
around $r$ fragments erased over time due to \nfail s, 
and the peak \rrepairrate\ is close to the average \rrepairrate.
The peak \rrepairrate\ for the \liqrepairer\ described in Section~\ref{liqrepairer sec}, 
is within a factor of two of the lower bounds on the \rrepairrate,
and the peak \rrepairrate\ for the \advliqrepairer\ described 
in Section~\ref{advliqrepairer sec} asymptotically approaches the lower bounds.

There are a number of possible strategies beyond those
outlined above that could be used to implement a distributed storage system.
One of our primary contributions is to provide fundamental lower bounds
on the \rrepairrate\ needed to maintain \srcdata\ recoverability for {\em any}
distributed storage system, current or future, 
for a given \storeoverhead\ and \nfail\ rate.

\section{Distributed storage model}

We introduce a model of distributed storage which is inspired by 
properties inherent and common to systems described in Section~\ref{practical sec}.
This model captures some of the essential features of any distributed
storage system.  All lower bounds are proved with respect to this model.

\subsection{Architecture}

\begin{figure}
\includegraphics[width=0.5\textwidth]{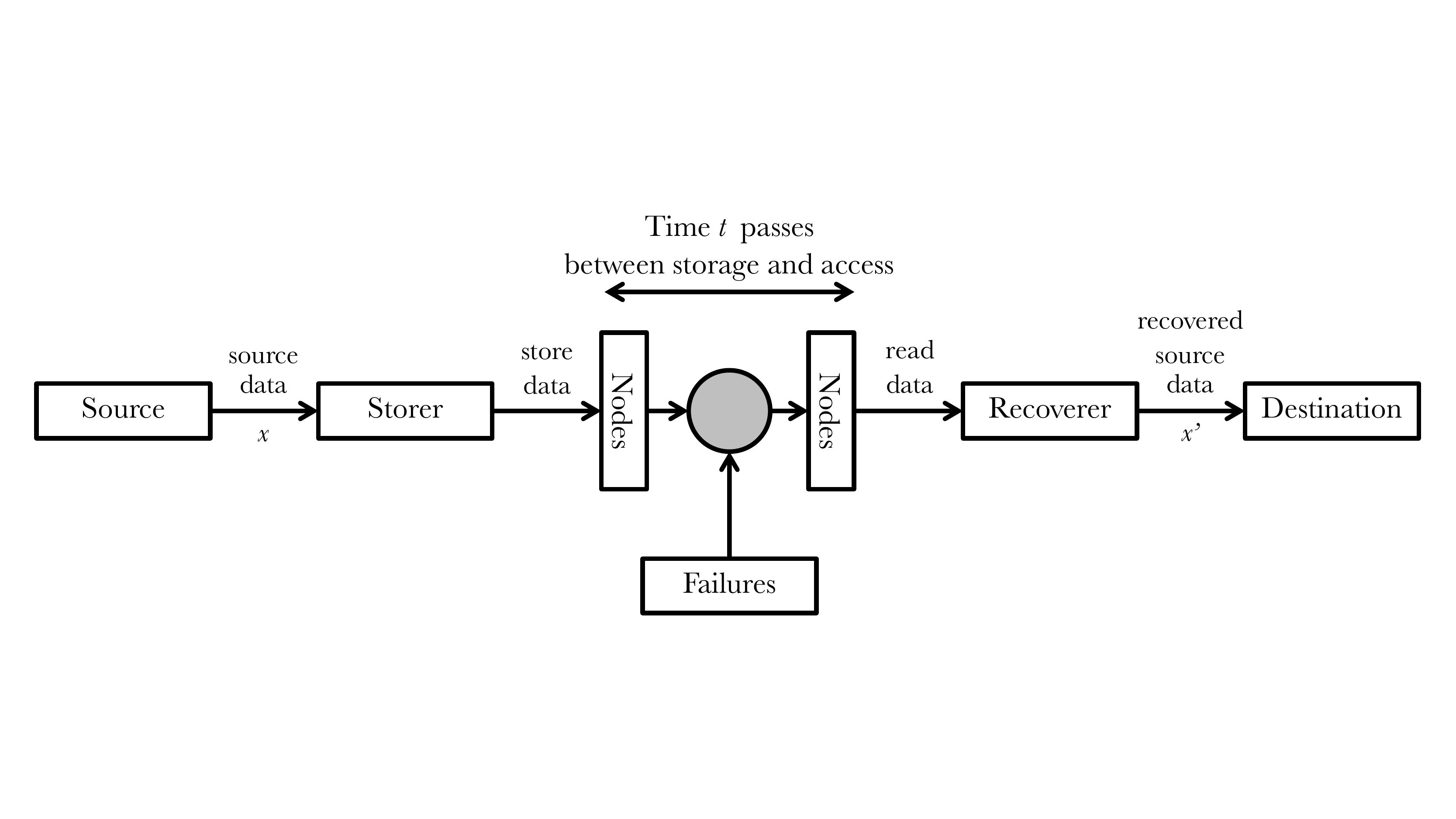}
\vspace{-0.7in}
\caption{Distributed storage architecture}
\label{storage_model fig}
\end{figure}

Figure~\ref{storage_model fig} shows an architectural overview of the distributed storage model.
A storer generates data from \srcdata\ $\xvar \in \{0,1\}^\xlen$ received from a source,
and stores the generated data at nodes.
In our model we assume the \srcdata\ is randomly and uniformly chosen,
and let random variable $\Xvar \inu \{0,1\}^\xlen$,
where $\inu$ indicates randomly and uniformly chosen.
Thus, $\len{\Xvar}= \entropy{\Xvar} =\xlen$, where
$\len{\Xvar}$ is the length of $\Xvar$ and $\entropy{\Xvar}$ 
is the entropy of $\Xvar$. 

\vspace{-0.15in}
\noindent

\begin{figure}
\includegraphics[width=0.50\textwidth]{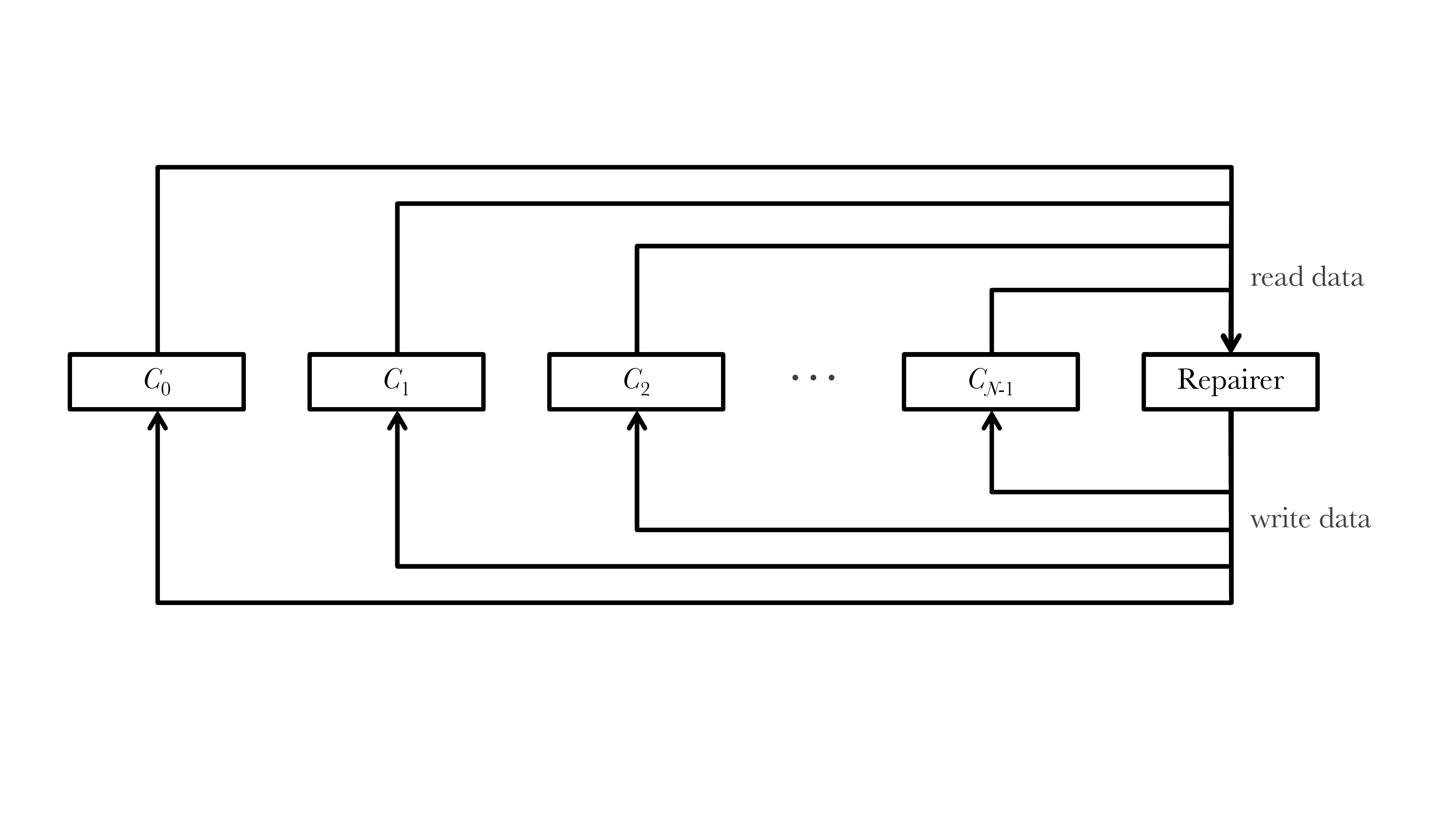}
\vspace{-0.7in}
\caption{Storage nodes and repairer model.}
\label{storage_nodes fig}
\end{figure}

Figure~\ref{storage_nodes fig} shows the nodes of the distributed storage system, 
together with the network that connects each node to a repairer.
Each of $\Ndata$ nodes $\Cdata_0,\ldots,\Cdata_{\Ndata-1}$ can store $\clen$ \bitval s,
and the {\em \rcapacity} is $\Ndata \cdot \clen$.
The {\em \storeoverhead} $\beta$ is the fraction of \rcapacity\ available beyond 
$\xlen$, i.e.,
\begin{equation}
\label{storage_over eq}
\beta = 1- \frac{\xlen}{\Ndata \cdot \clen},
\end{equation}
thus $\xlen = (1-\beta) \cdot \Ndata \cdot \clen$.

\vspace{-0.15in}
\noindent

As nodes fail and are replaced, a repairer continually reads data from the nodes, 
computes a function of the read data, and writes the computed data back to the nodes.  
The repairer tries to ensure that the \srcdata\ can be recovered at any time
from the data stored at the nodes.

As shown in Figure~\ref{storage_model fig},
after some amount of time $\tvar$ passes, a recoverer reads data from the nodes to
generate $\xvar'$, which is provided to a destination, where $\xvar$ 
is reliably recovered if $\xvar' = \xvar$.
The goal is to maximize the amount of time $\tvar$ the recoverer can reliably recover $\xvar$.
  
\subsection{\Nfail s}

A \nfseq\ determines when and what nodes fail as time passes.
A \nfseq\ is a combination of two sequences,
a {\em \timeseq}
 \[\tvar_{0} \le \tvar_{1} \le \cdots 
 \le\tvar_{i} \cdots, \]
where for {\em index} $i$,  $\tvar_{i}$ is the {\em time} at which a node fails, and an {\em \idseq}
\[\yvar_0,\yvar_{1},\ldots,\yvar_{i},\ldots,\]
where $\yvar_{i}$ is the {\em \identifier} of the node that fails at time $\tvar_{i}$.

All $\clen$ \bitval s stored at node $\yvar_{i}$ are 
immediately erased at time $\tvar_{i}$ when the node fails, 
i.e., all $\clen$ \bitval s of node $\yvar_{i}$ are
initialized to zero at time $\tvar_{i}$ when the node
fails.  This can be viewed as immediately replacing 
a failed node with a replacement node 
with storage initialized to zeroes.
Thus, at each time there are $\Ndata$ nodes.

A primary objective of practical distributed storage architectures is to distribute the
components of the system so that failures are as independent as possible.
{\em \Ptdist s} are an idealization of this primary objective, 
and are often used to model and evaluate distributed storage systems in practice.
For a \Ptdist\ with rate $\lambda$, the time between when a node is 
initialized and when it fails is an exponential random variable with rate $\lambda$,
i.e., $\frac{1}{\lambda}$ is the average lifetime of a node between
when it is initialized and when it fails.
Our main lower bounds in Section~\ref{main sec} 
are with respect to \Ptdist s.

\subsection{Network}
\label{network sec}

The model assumes there is a network interface between 
each node and the system over which all data from and to the node travels.
One of the primary lower bound metrics is the amount of data
that travels over interfaces from nodes to the system, which is counted
as data read by the system.   For the lower bounds, this is the only
network traffic that is counted. All other data traffic within
the system, i.e. data traffic across the system,
data traffic over an interface from the system to nodes,
or any other data traffic that does not travel over an interface 
from a node to the system, is not counted for the lower bounds.
It is assumed that the network is completely reliable, and that
all data that travels over an interface from a node to the system
is instantly available everywhere within the system.   

\subsection{Storer}
\label{storer sec}
A storer takes the \srcdata\ $\xvar$ and generates and
stores data at the nodes in a preprocessing step 
when the system is first initialized and before there are any \nfail s.
We assume that the recoverer can reliably recover $\xvar$ 
from the data stored at the nodes 
immediately after the preprocessing step finishes. 

For simplicity, we view the storer preprocessing step as part of the repairer, 
and any data read during the storer preprocessing step is not counted in the lower bounds.

For the lower bounds, there are no assumptions about 
how the storer generates the stored data from the \srcdata,
i.e. no assumptions about any type of coding used, no assumptions
about partitioning the \srcdata\ into objects, etc.
As an example, the \srcdata\ can be encrypted, compressed,
encoded using an error-correcting code or erasure code, replicated, 
or processed in any other way known or unknown to generate
the stored data, and still the lower bounds hold.  Analogous
remarks hold for the repairers described next.

\subsection{Repairer}
\label{repairer sec}
 
A {\em repairer} ensures that the \srcdata\ is recoverable 
when data generated from the \srcdata\ is stored at the unreliable nodes.
A repairer for a system operates as follows.
The \identifier\ $\yvar_{i}$ is provided to a repairer
at time $\tvar_{i}$, which alerts the repairer that all $\clen$
\bitval s stored on node $\yvar_{i}$ are lost at that time.
As nodes fail and are replaced, the repairer reads data over interfaces from nodes, 
performs computations on the read data, and writes computed data
over interfaces to nodes.
A primary metric is the number of bits the repairer reads over interfaces
from storage nodes.

Let $\Lseq(\tvar)  = (\tseq,\yseq)$ be the \nfseq\ 
up till time $\tvar$, where 
$\tseq = \{\tvar_{0},\ldots,\tvar_{\ell}\},$
$\yseq = \{\yvar_{0},\ldots,\yvar_{\ell}\},$
and $\ell = \argmax_i \{\tvar_i \le t \}$.
The repairer $\Rfnc$ has access to $\Lseq(\tvar)$ at time $\tvar$ at no cost.

Let
$\Vdata(\tvar)$ be the \bitval s stored in the global memory of $\Rfnc$, 
where $\vlen = \len{\Vdata(\tvar)}$,
\[ \Cdata_0(\tvar),\ldots,\Cdata_{\Ndata-1}(\tvar) \]
be the \bitval s stored at nodes $0,\ldots,\Ndata-1$,
respectively, where $\clen = \len{\Cdata_j(\tvar)}$ is the 
capacity of each node $j$, and
\[ \Sdata(\tvar) = \{\Vdata(\tvar), 
\{ \Cdata_0(\tvar),\ldots,\Cdata_{\Ndata-1}(\tvar) \} \} \]
is the global state of the system at time $\tvar$, 
where $\slen = \len{\Sdata(\tvar)}$.
Thus, 
\[  \slen = \vlen + \Ndata \cdot \clen \]
is the size of the global system state at any time $\tvar$.

The actions of $\Rfnc$ at time $\tvar$ are 
determined by $(\tvar,\Vdata(\tvar),\Lseq(\tvar))$.  
If a node $j$ fails at time $\tvar$ then $\Rfnc$ is notified at time $\tvar$ 
that node $j$ failed and $\Lseq(\tvar)$ is updated.  If $\Rfnc$ reads data over 
the interface from node $j$ at time $\tvar$ (when to read from node $j$ 
is determined by $(\tvar,\Vdata(\tvar),\Lseq(\tvar))$)
then the amount and location of the data read from $\Cdata_j(\tvar)$ 
is determined by $(\tvar,\Vdata(\tvar),\Lseq(\tvar))$.
The data read over the interface from node $j$ in response to a request
initiated at time $\tvar$ is assumed to be instantaneously available, 
i.e. all of the requested data is available at time $\tvar$ 
over the interface from node $j$.

$\Vdata(\tvar)$ can be used by the repairer to store 
the programs the repairer executes, store information 
from the past, temporarily store data read from nodes, 
perform computations on read data, temporarily
store computed data before it is written to nodes, and generally to store any
information the repairer needs immediate access to that is not stored at the nodes.  
The distinction between $\Vdata(\tvar)$ and the nodes 
is that $\Vdata(\tvar)$ is persistent memory
 (not subject to any type of failure in the model) and available globally to 
 $\Rfnc$ (there is no read or write cost for accessing $\Vdata(\tvar)$).
 $\Rfnc$ can also store such information at the nodes, but this information
 is subject to loss due to possible \nfail s.

Repairers are allowed to use an unbounded amount of computation,
since computation time is not a metric of interest in the lower bounds.
The granularity of how much data is read or written 
in one step is unconstrained, e.g. one bit or Terrabytes
of data can be read over an interface from a node during a read step,
and the lower bounds still hold. The granularity of the timing
of read and write steps is also unconstrained, e.g. there may be
a read step each nanosecond, or every twenty minutes. 

{\em \Lrepairer s}, inspired by \cite{Dimakis07} 
and \cite{Dimakis10}, are more powerful than repairers.
The motivation for the \lrepairer\ model
is that a node often has CPUs, memory and storage, 
and often the impact of traffic between storage and memory
at a node is much less than the impact of traffic over the interface
from the node to the system.  Thus, an arbitrary amount of data
may be accessed locally from storage into local memory at a node, local CPUs
may compute and store in the local memory a much smaller amount of data
from the data accessed into local memory, and it is the much smaller amount of
data computed by the CPUs that is sent over the interface from the node to the system. 
The model does not count the data accessed from storage into local memory,
it only counts the data in the local memory that is read by the 
system over the interface from the node.

Formally, for a \lrepairer, when data is to be read over the interface 
from node $j$ initiated at time $\tvar$ (when to read from node $j$ is determined 
by $(\tvar,\Vdata(\tvar),\Lseq(\tvar))$), a copy of the entire global memory of the 
\lrepairer\ is assumed to be instantaneously available
 in the local memory at node $j$ at no cost.  As the local computation
 at node $j$ progresses, the copy may evolve to be different than the global
 memory of the \lrepairer\ at time $\tvar$, 
 but the only information the \lrepairer\ potentially receives 
 about any changes to the copy in the local memory 
 is from the locally computed data read 
 by the \lrepairer\ over the interface from node $j$. 
The locally computed data is generated 
based on $(\tvar,\Vdata(\tvar),\Lseq(\tvar),\Cdata_j(\tvar))$, 
and then the locally computed data is read by the 
\lrepairer\ over the interface from node $j$.  
The local computational power at node $j$ and the throughput of the interface
at node $j$ are assumed to be unlimited,
and thus the locally computed data requested at time $\tvar$ by the \lrepairer\ is 
available instantly at time $\tvar$ over the interface from node $j$.

Thus the data read over the interface from node $j$ when the request for
the data is initiated at time $\tvar$ is determined by 
$(\tvar,\Vdata(\tvar),\Lseq(\tvar),\Cdata_j(\tvar))$.
In this model only the locally computed data is counted 
as data read over the interface from node $j$;  
the data accessed from storage at node $j$ to produce the locally
computed data (which could be all of $\Cdata_j(\tvar)$) is not counted. 

For example, in the extreme a \lrepairer\ could locally access 
all data stored at a node to produce $1$ KB of locally computed data, and 
then only the $1$ KB of locally computed data is read over the interface from
the node.  In this example, only $1$ KB of data is counted towards  
data read by the \lrepairer.  
Thus there is a significant cost to this generalization 
that is not counted in the amount of 
data read from nodes by the repairer.  
These issues are discussed in more detail in Section~\ref{related work sec}.

A repairer is a special case of a \lrepairer: a repairer 
is simply a \lrepairer\ where the data accessed from storage at
the node is directly sent over the interface from the node to the repairer. 

A repairer may employ a randomized algorithm,
which could be modeled by augmenting the repairer
with random and independently chosen \bitval s.   
However, since the repairer is deterministic for 
a fixed setting of the random bits and the lower bounds 
hold for any deterministic repairer, the same lower bounds
hold for any randomized repairer.
Thus, we describe lower bounds only for deterministic repairers, noting that all the lower bound results immediately carry over to randomized repairers. 

\subsection{Recoverer}
\label{recovery sec}

For any repairer $\Rfnc$ there is a recoverer $\Afnc$ 
such that if the \srcdata\ is $\xvar$ and the state at time $\tvar$ is $\Sdata(\tvar)$ when the repairer is $\Rfnc$ 
then $\Afnc(\Sdata(\tvar))$ should be equal to $\xvar$.

\Srcdata\ $x$ is {\em recoverable} at time $\tvar$
with respect to repairer $\Rfnc$ and recoverer $\Afnc$ if
$\Afnc(\Sdata(\tvar)) = \xvar$.
\Srcdata\ $x$ is {\em unrecoverable} at time $\tvar$
with respect to repairer $\Rfnc$ and recoverer $\Afnc$ if
$\Afnc(\Sdata(\tvar)) \not= \xvar$.

\subsection{Applying the lower bounds to real systems}
\label{real system sec}

The description of the model makes some very unrealistic assumptions
about how real systems operate in practice.   
However, it is these assumptions that ensure that 
the lower bounds apply to all real systems.
Consider a real system where nodes fail randomly as in the model,
but also portions of the network intermittently fail,
network bandwidth availability is limited and varying
between different parts of the system, 
memory is not completely reliable, multiple distributed semi-autonomous
processes are interacting with different sets of nodes,
responses are not immediate to data requests over
node interfaces,
processes are not immediately notified when nodes fail, 
notification of node failure is not global, 
nodes are not immediately replaced, computational resources are limited, etc.
We describe an omniscient agent acting 
with respect to the model in the role
of the repairer, where the agent emulates the processes
and behaviors of the real system.  
This shows the lower bounds also apply to the real system.

In the model, nodes that fail are immediately 
replaced and the agent is immediately notified.
In the real system, a failed node may not be replaced immediately.
Thus, to emulate the real system, the agent disallows any response to a request to read or write data 
to a failed node from a process until the time 
when the node would have been replaced in the real system. 

In the real system, notifications of node failures may not be instantaneous, and only some processes may be notified.
Thus, the agent only notifies the appropriate processes  
of node failures when they would have been notified in the real system.  

In the model, the agent receives an immediate and complete
response to a request for data over an interface to a node. 
In the real system, interfaces can have a limited amount of 
bandwidth, and there can be delays in delivering responses 
to requests for data by processes over a node interface
due to computational limits or other constraints.
Thus, the agent delivers data to requesting processes 
with the delays and at the speed of the real system.

In the real system, only a small portion of the global
memory state may be available in the local memory
of a node when local-computation repair is used.
Thus, the agent may only need a small portion of the
global memory at the node to emulate a \lrepairer\
of the real system.  

In the model, the agent acting as a repairer has one global memory.
In the real system, repair may be implemented by
a distributed set of processes $\Rfnc_1,\ldots, \Rfnc_i$
executing concurrent reads and writes over node interfaces,
each with their own private memory 
$\Vdata_1(\tvar), \ldots,\Vdata_i(\tvar)$ at time $\tvar$.  
The agent $\Rfnc$ can emulate $\Rfnc_1,\ldots, \Rfnc_i$ as follows.
The global memory of $\Rfnc$ is
\[\Vdata(\tvar) = \{\Vdata_1(\tvar), \ldots,\Vdata_i(\tvar) \}. \] 
If processes $\Rfnc_i$ and $\Rfnc_j$ send \bitval s 
between their local memories at time $\tvar$ then 
these same \bitval s are copied between $\Vdata_i(\tvar)$ 
and $\Vdata_j(\tvar)$ by $\Rfnc$ at time $\tvar$. 
The movement of data between the local memories 
of the processes that the agent is emulating is at no cost.
Thus, the lower bound on the amount of data read over interfaces from 
nodes by $\Rfnc$ in the model is a lower bound on the amount of data 
read over interfaces from nodes by $\Rfnc_1,\ldots, \Rfnc_i$.  

In the model, the agent has a single interface with each node.   
In the real system, a node can have multiple interfaces.
These multiple interfaces are considered as one logical interface by the agent
when counting the amount of data traveling over interfaces from nodes to the agent,
and the agent delivers data traveling over the multiple interfaces to the 
appropriate requesting processes of the emulated real system.

The count of data traffic for the lower bounds is conservative, 
i.e. the amount of data that travels over interfaces from nodes to the agent is a lower bound on the amount 
of data traveling over the network in the real system.

Thus, a real system, whether it is
perfectly architected and has non-failing infinite
network bandwidth, zero computational delays,
 instant node failure notification, or whether it
is more realistic as described above, can be emulated
by the agent in the model as described above. 
Since the lower bounds apply to the agent with respect to the 
model, the lower bounds also apply to any real system.

\subsection{Practical parameters example}
\label{practical setting sec}
A practical system can have $\Ndata = \expp{5}$ nodes, 
$\clen = \expp{16}$ \bitval s of capacity at each node, 
thus $\Ndata \cdot \clen = \expp{21}$.
The amount of storage needed by the repairer to store its programs and state 
generously is at most something like $\vlen = \expp{13}$ \bitval s.
Generally, $\clen >> \vlen >> \Ndata$. 
We assume $\clen \ge \Ndata$ and $\vlen << \Ndata \cdot \clen$
in our bounds with respect to growing $\Ndata$.

Practical values of $\beta$ range from $2/3$ (triplication) to $1/20$ and smaller.
In the example, $\xlen = (1-\beta) \cdot \Ndata \cdot \clen \approx \expp{21}$.
In practice nodes fail each few years, e.g., $1/\lambda = $ 3 years.

\section{Emulating repairers in phases}
\label{universal sec}

We prove lower bounds based on considering 
the actions of a repairer $\Rfnc$, or \lrepairer\ $\Rfnc$,
running in phases on a \nfseq.  Each phase considers
a portion of a \nfseq\ with $M$ \nfail s, where each
of the $M$ \nfail s within a phase are distinct, as described in more detail below.

For any $M \le \Ndata$, we write 
\[ \ang{\yvar_0,\ldots,\yvar_{M-1}} \]
when all \identifier s are distinct, i.e., $\yvar_{i} \not= \yvar_{i'}$
for $0 \le i \not= i' \le M-1$.
We write 
\[ \ang{\yvar_0,\ldots,\yvar_{j-1}, \Yvar_j,\ldots,\Yvar_{M-1}} \]
when $\ang{\yvar_0,\ldots,\yvar_{j-1}}$ are distinct \identifier s,
random variable $\Yvar_j$ is defined as
\[ \Yvar_{j} \inu \{0,\ldots,\Ndata-1\} - \{\yvar_0,\ldots,\yvar_{j-1}\}, \]
and for $i =j+1,\ldots,M-1$, random variable $\Yvar_i$ is defined as
\[ \Yvar_{i} \inu \{0,\ldots,\Ndata-1\} - \{\yvar_0,\ldots,\yvar_{j-1},\Yvar_j,\ldots\Yvar_{i-1}\}, \]
where $\inu$ indicates randomly and uniformly chosen.
Thus, $\ang{\yvar_0,\ldots,\yvar_{j-1}, \Yvar_j,\ldots,\Yvar_{M-1}}$ is a
distribution on distinct \identifier s.

A phase consists of executing $\Rfnc$ on a portion of a \nfseq\ $(\tseq,\yseq)$, where 
\[ \tseq = \{\tvar_{0}, \tvar_{1} , \ldots ,\tvar_{M-1} \} \]
is the portion of the \timeseq\ and
\[ \yseq = \ang{\yvar_0, \yvar_{1}, \ldots, \yvar_{M-1}} \]
is the portion of the \idseq\ that is revealed to $\Rfnc$ 
as the phase progresses.

Before a phase begins, the storer generated and stored data at the nodes
based on \srcdata\ $\xvar$, and the repairer $\Rfnc$ has been
executed with respect to a \nfseq\ up till time $\tvarm_0$, 
where $\tvarm_0$ is just before 
the time of the first \nfail\ of the phase at time $\tvar_{0}$.  
We assume that the recoverer $\Afnc$ can recover \srcdata\ $\xvar$ 
from the state $\Sdata(\tvarm_0)$.

Conceptually, there are two executions of $\Rfnc$ on $(\tseq, \yseq)$ in a phase.
The first execution runs $\Rfnc$ normally from $\tvarm_0$ to $\tvarp_{M-1}$
starting system state $\Sdata(\tvarm_0)$ and ending in $\Sdata(\tvarp_{M-1})$,
where $\tvarm_0$ is just before $\tvar_{0}$, 
and $\tvarp_{M-1}$ is just after $\tvar_{M-1}$.
Thus, the \nfail s at times $\tvar_{0}$ and $\tvar_{M-1}$
are within the phase, but $\Rfnc$ does not read 
any \bitval s before $\tvar_{0}$ or after $\tvar_{M-1}$ in the phase.

The following compressed state is defined by the first execution of $\Rfnc$. 
For $i = 1,\ldots,M-1$,  let $\Rdata_{\yvar_i}$ be the
concatenation of the \bitval s read by 
repairer $\Rfnc$ over the interface from node $\yvar_i$
between $\tvar_{0}$ and $\tvar_i$.
More generally, if $\Rfnc$ is a \lrepairer, then $\Rdata_{\yvar_i}$ is the 
concatenation of the locally-computed
bits read by $\Rfnc$ over the interface
from node $\yvar_i$ between $\tvar_{0}$ and $\tvar_i$.
Let
\begin{equation*}
\Ddata = \{ \Vdata(\tvarm_0), 
\{ \Cdata_j(\tvarm_0): j \not\in \yseq\},  
\{ \Rdata_j: j \in \yseq \}\}
\end{equation*}
be the {\em compressed state} with respect to $(\xvar,(\tseq,\yseq))$. 
 The first execution of $\Rfnc$ shows
that $\Ddata$ can be generated based on repairer $\Rfnc$, $\Sdata(\tvarm_0)$ 
and $(\tseq,\yseq)$.

The second execution uses the compressed state $\Ddata$ in place of $\Sdata(\tvarm_0)$
to emulate $\Rfnc$ on $(\tseq,\yseq)$ and arrive in the same final state 
$\Sdata(\tvarp_{M-1})$ as the first execution. 
The initial memory state of $\Rfnc$ is set to $\Vdata(\tvarm_0)$
and the state of node $j$ is initialized to $\Cdata_j(\tvarm_0)$ for all $j \notin \yseq$.   
Function $f:\{0,\ldots,\Ndata-1\} \rightarrow \{0,1\}$ is initialized as follows: 
$f(j) = 1$ if $j \in \yseq$ and $f(j) = 0$ if $j \not\in \yseq$.
$\Rfnc$ is emulated from $\tvarm_0$ to $\tvarp_{M-1}$ the same as in
the first execution with the following differences.
When $\Rfnc$ is to read \bitval s over the interface 
from node $j$ at time $\tvar$:  if $f(j) == 0$ then
the requested \bitval s are read from $S_j(\tvar)$; 
if $f(j) == 1$ then the requested \bitval s are provided to $\Rfnc$ 
from the next portion of $\Rdata_j$ not yet provided to $\Rfnc$. 
When $\Rfnc$ is to write \bitval s to a node $j$ at time $\tvar$:
if $f(j) == 0$ then the \bitval s are written to $S_j(\tvar)$;
if $f(j) == 1$ then the write is skipped.  
At time $\tvar$ when $j \in \yseq$ first fails within the phase, 
$f(j)$ is reset to $0$ and $S_j(\tvar)$ is initialized to all zero \bitval s.  

If $\Rfnc$ is a \lrepairer\ instead of a repairer then
when $\Rfnc$ is to produce and read locally-computed \bitval s 
over the interface from node $j$ at time $\tvar$ and $f(j) == 0$ 
the requested bits are locally-computed
by $\Rfnc$ based also on $\Cdata_j(\tvar)$. 

It can be verified that the state of the system is 
$\Sdata(\tvarp_{M-1})$ at the end of the emulation,
whether $\Rfnc$ is a repairer or a \lrepairer.
Thus, $(\Sdata(\tvarp_{M-1}),(\tseq,\yseq))$ can be generated from
$(\Ddata,(\tseq,\yseq))$ based on $\Rfnc$.

A key intuition is that if repairer $\Rfnc$ doesn't read enough data 
over interfaces from nodes that fail before they fail during a phase
then $\len{\Ddata} < \xlen$ and thus $\xvar$ cannot
be reliably recovered from $\Ddata$.  On the other hand, $\xvar$
is recoverable at time $\tvarp_{M-1}$ by $\Afnc$ only if $\xvar$ can be
recovered from $\Ddata$ since $\Ddata$ can generate $\Sdata(\tvarp_{M-1})$
and $\Afnc(\Sdata(\tvarp_{M-1}))$ is supposed to equal $\xvar$.  
Thus, if repairer $\Rfnc$ doesn't read enough data over interfaces
from nodes that fail before they fail during a phase then $\xvar$
is unrecoverable at time $\tvarp_{M-1}$ by $\Afnc$.  
We formalize this intuition below.

We let 
\[ \xvar \rightarrow_{(\tseq,\yseq)} \Ddata \rightarrow_{(\tseq,\yseq)} \xvar' \] 
indicate that \srcdata\ $\xvar$ is mapped before the start of the phase to
a value of $\Sdata(\tvarm_0)$ by $\Rfnc$, which in turn is mapped by $(\tseq,\yseq)$
to a value $\Ddata$ by the first execution of the emulation of $\Rfnc$, 
which is mapped to a value of 
$\Sdata(\tvarp_{M-1})$ by $(\tseq,\yseq)$ by the second execution 
of the emulation of $\Rfnc$, which in turn is mapped 
by $\Afnc$ to a value $\xvar' \in \{0,1\}^\xlen$.

\vspace{0.1in}
\begin{complemma}
\label{compression lemma}
For any repairer or \lrepairer\ $\Rfnc$ and recoverer $\Afnc$,
for any $(\tseq,\yseq)$, for any 
\begin{equation*}
\Xset \subseteq \Xset_{(\tseq,\yseq)} = 
\left\{ \xvar: \xvar \rightarrow_{(\tseq,\yseq)}  \Ddata \in \{0,1\}^{\xlen-\ell} \right\},
\end{equation*}
\begin{gather*}
\Probi{\Xvar,(\tseq,\yseq)}{(\Xvar \in \Xset) \wedge 
(\Xvar \mbox{ \rm recoverable at } \tvarp_{M-1})} \\
\le 2^{-\ell}. \nonumber
\end{gather*}
\end{complemma}
\begin{proof}
For any $(\tseq,\yseq)$, for any $\Xset \subseteq \Xset_{(\tseq,\yseq)}$,
for any $\xvar \in \Xset$,
\[ \xvar \rightarrow_{(\tseq,\yseq)} \Ddata \rightarrow_{(\tseq,\yseq)} \xvar' \]
for some $\Ddata \in \{0,1\}^{\xlen-\ell}$ and some $\xvar' \in \{0,1\}^\xlen$.
Since $\Ddata \in \{0,1\}^{\xlen-\ell}$, 
the number of possible $x'$ values is at most $2^{\xlen-\ell}$,
thus $\xvar = \xvar'$ for at most $2^{\xlen-\ell}$ 
of the $\xvar \in \Xset$.
\qed
\end{proof}

Let
\begin{equation}
\label{ulen eq}
\ulen = \sum_{j\in \yseq} \len{R_{j}}
\end{equation}
with respect to $(\xvar,(\tseq,\yseq))$
be the total length of data read by $\Rfnc$
from nodes that fail before their failure
in the phase.
Then,
\begin{equation}
\label{sddata eq}
\len{\Ddata} = \vlen + (\Ndata - M) \cdot \clen+ \ulen
\end{equation}
with respect to $(\xvar,(\tseq,\yseq))$.
Let
\begin{equation}
\label{sfdata eq}
\olen = \Ndata \cdot \clen - \xlen + \vlen +1,
\end{equation}
and let 
\begin{equation}
\label{fdata eq}
\Fdata = \left\lceil \frac{\olen}{s} \right\rceil
\end{equation}
be the minimal number of nodes so that 
$\Fdata \cdot \clen \ge \olen$.
Let
\begin{equation}
\label{betap eq}
\betap = \frac{\Fdata}{\Ndata}
\end{equation}
Note that 
\[ \beta \le \betap \le  
\beta+\frac{\vlen+1}{\Ndata \cdot \clen} + \frac{1}{\Ndata}. \] 
Generally, $\betap \approx \beta$, e.g., for the practical system 
described in Subsection~\ref{practical setting sec}, 
$\betap \le \beta + \expm{7} + \expm{4}$.

Hereafter we set $M = 2 \cdot \Fdata < \Ndata$, which implies  $\betap < 1/2$.
This restriction is mild 
since $\betap \rightarrow 0$ is more interesting in practice than $\betap \approx 1$.

\vspace{0.1in}
\begin{compcorollary}
\label{compression corollary}
For any repairer or \lrepairer\ $\Rfnc$ and recoverer $\Afnc$,
for any $(\tseq,\yseq)$, for any 
\begin{equation*}
\Xset \subseteq \Xset_{(\tseq,\yseq)} =
\left\{ \xvar: \ulen \le (\Fdata-1) \cdot \clen \right\}
\end{equation*}
where $\ulen$ is defined with respect to $(\xvar,(\tseq,\yseq))$,
\begin{gather}
\label{compression corollary eq}
\Probi{\Xvar,(\tseq,\yseq)}{(\Xvar \in \Xset) \wedge 
(\Xvar \mbox{ \rm recoverable at } \tvarp_{M-1})} \\
\le 2^{-\clen}. \nonumber
\end{gather}
\end{compcorollary}
\begin{proof}
Follows from \Complemma\ 
with $M = 2 \cdot \Fdata \le \Ndata$,
$\ell = \clen$, and using Equations~\eqref{sddata eq}, \eqref{sfdata eq}, \eqref{fdata eq}.
\qed
\end{proof}
Note that $2^{-\clen}$ is essentially zero in any practical setting.  For example,
$2^{-\clen}\le 10^{-3 \cdot 10^{15}}$ for the settings in Section~\ref{practical setting sec}.

Note that \Complemma\ and \Compcorollary\ rely upon the
assumption that the \srcdata\ is uniformly distributed, 
and this is the root of the dependency of all 
subsequent technical results on this assumption.

A natural relaxation of this assumption is that the
\srcdata\ has high min-entropy, where the min-entropy
is the log base two of one over the probability of $\xvar$, 
where $\xvar$ is the most likely value for the \srcdata.
Thus the min-entropy of the \srcdata\ is always at most $\xlen$.

Since the \srcdata\ for practical systems is composed
of many independent source objects, typically the min-entropy
of the \srcdata\ for a practical system is close to $\xlen$.
It can be verified that all of the lower bounds hold if 
the min-entropy of the \srcdata\ is universally substituted for $\xlen$.

\section{Core lower bounds}
\label{core lower sec}

From \Compcorollary,  a necessary condition
for \srcdata\ $\xvar$ to be reliably recoverable at the end of the phase
is that repairer or \lrepairer\ $\Rfnc$ must read a lot of data from nodes that fail during the phase, 
and $\Rfnc$ must read this data before the nodes fail.  

On the other hand, $\Rfnc$ cannot predict which nodes are going to fail
during a phase, and only a small fraction of the nodes fail during a phase.
Thus,  to ensure that enough data has been read from nodes 
that fail before the end of the phase, a larger 
amount of data must be read in aggregate from all the nodes.

\Corelemma, the primary technical contribution of this section,
is used to prove \Coretheorem\ and all later results.

Fix $\xvar$, $\tseq = \{\tvar_0,\ldots,\tvar_{M-1}\}$,  
and $\yseq = \ang{\yvar_0,\ldots,\yvar_{M-1}}$. 

For $i \in \{0,\ldots,M-1\}$, let 
$\tseq_i = \{\tvar_0,\ldots,\tvar_{i}\}$ 
be a prefix of $\tseq$, and let 
$\yseq_i = \ang{\yvar_0,\ldots,\yvar_{i}}$ 
be a prefix of $\yseq$.  
For $i \in \{0,\ldots,M-1\}$, $j \in \{0,\ldots,\Ndata-1\}$, let 
\[ \rlen_{i,j} \]
with respect to $(\xvar, (\tseq_i,\yseq_{i-1}))$
be the amount of data read from node $j$ 
up till time $\tvar_{i}$.  
For $i \in \{0,\ldots,M-1\}$, let
\[ \rlen_{i} = \sum_{j \in \{0,\ldots,\Ndata-1\}} \rlen_{i,j} \]
with respect to $(\xvar, (\tseq_i,\yseq_{i-1}))$.
Let
\[ \ulen_{i}= \rlen_{i,\yvar_{i}} \]
with respect to $(\xvar, (\tseq_i,\yseq_i))$
be the total number of \bitval s read from the node $\yvar_i$ up till it fails at time $\tvar_{i}$. 
Then,
\begin{equation*}
\ulen = \sum_{i=1}^{M-1} \ulen_{i}
\end{equation*}
with respect to $(\xvar,(\tseq,\yseq))$, 
where $\ulen$ is defined in Equation~\eqref{ulen eq}.

\vspace{0.1in}
\begin{corelemma}
\label{core lemma}
Fix $\betap < 1/2$. Fix $\epscore > 0$
and let 
\begin{equation}
\label{delcore eq}
\delcore =  2 \cdot F \cdot e^{-\frac{\epscore^2 \cdot \Fdata}{4}+\epscore}.
\end{equation}
For $i = 1, \dots, 2 \cdot \Fdata-1$, let 
\begin{equation}
\label{gamma eq}
\Gamma_i = (1-\epscore) \cdot \frac{i \cdot \left(\Ndata-\frac{i+1}{2}\right) \cdot \clen}
{2 \cdot \Fdata -1}.
\end{equation}
For any repairer or \lrepairer, 
\begin{gather*}
\nonumber
 \Probi{\xvar,(\tseq,\Yseq)}{\left(\forall_{i=1}^{2 \cdot \Fdata -1}\rlen_{i} < 
\Gamma_i \right) \wedge (\ulen > (\Fdata-1) \cdot \clen)} \\
 \le \delcore, 
\end{gather*}
for any $\xvar$ , $\tseq$, and 
$\Yseq = \ang{\yvar_0, \Yvar_1\ldots,\Yvar_{2 \cdot \Fdata -1}}$ for any $\yvar_0$.
\end{corelemma}
\begin{proof}
Fix $\xvar$, $\tseq$ and $\yvar_0$. The parameterization with respect to
$\xvar$ and $\tseq$ are implicit in the remainder of the proof. 
Fix $\eta = (\Fdata-1) \cdot \clen$.
We first prove that for any repairer or \lrepairer\ $\Rpfnc$ there is a repairer or \lrepairer\ $\Rfnc$ such that
\begin{gather}
\label{repairer reduc eq}
\Prob{\left(\forall_{i=1}^{2 \cdot \Fdata -1}\rlen'_{i} < 
\Gamma_i \right) \wedge (\ulen' > \eta)} \\
\le \Prob{\ulen > \eta} \nonumber
\end{gather}
and
\begin{equation}
\label{repairer less eq}
\Prob{\forall_{i=1}^{2 \cdot \Fdata -1}\rlen_{i}< \Gamma_i } = 1 
\end{equation}
with respect to $\ang{\yvar_0, \Yvar_1\ldots,\Yvar_{2 \cdot \Fdata -1}}$,
and where $\rlen'_i$ and $\ulen'$ are defined with respect to $\Rpfnc$ and
$\rlen_i$ and $\ulen$ are defined with respect to $\Rfnc$.

Let predicate $\Ppred$ be defined as follows on input 
$\ang{\yvar_0,\ldots,\yvar_{M-1}}$.
\begin{equation*}
\Ppred \mbox{ \rm is true }  \iff \forall_{i=1}^{2 \cdot \Fdata -1}\rlen'_{i}< \Gamma_i 
\end{equation*} 
with respect to $\ang{\yvar_0,\ldots,\yvar_{M-1}}$.

$\Rfnc$ acts the same way as $\Rpfnc$
with respect to  $\ang{\yvar_0,\ldots,\yvar_{M-1}}$ 
for which $\Ppred$ is true, thus
$\rlen_i = \rlen'_i$ for $i = 1,\ldots, M-1$, and $\ulen = \ulen'$,
with respect to  $\ang{\yvar_0,\ldots,\yvar_{M-1}}$
for which $\Ppred$ is true.

Fix  $\ang{\yvar_0,\ldots,\yvar_{M-1}}$ for which $\Ppred$ is false, let 
\[ \ell = \argmin_{i=1,\ldots,M-1} \{ \rlen'_i \ge \Gamma_i \} \]
with respect to  $\ang{\yvar_0,\ldots,\yvar_{M-1}}$.
$\Rfnc$ acts the same way up till time $\tvar_{\ell-1}$, but 
doesn't read data from nodes after $\tvar_{\ell-1}$,
with respect to  $\ang{\yvar_0,\ldots,\yvar_{M-1}}$.
Thus, $\rlen_i = \rlen'_i$ for $i = 1,\ldots, \ell-1$,
 $\rlen_i = \rlen'_{\ell-1}$ for $i = \ell,\ldots,M-1$,
with respect to $\ang{\yvar_0,\ldots,\yvar_{M-1}}$.
From this, $\forall_{i=1}^{2 \cdot \Fdata -1}\rlen_{i}< \Gamma_i$
with respect to any $\ang{\yvar_0,\ldots,\yvar_{M-1}}$
for which $\Ppred$ is false.  Thus,
condition~\eqref{repairer less eq} holds for repairer $\Rfnc$.

It can also be verified that $\ulen = \ulen'$ with respect to  
all  $\ang{\yvar_0,\ldots,\yvar_{M-1}}$ for which $\Ppred$ is true.
From this it follows that
\begin{gather*}
\Prob{\left(\forall_{i=1}^{2 \cdot \Fdata -1}\rlen'_{i} < 
\Gamma_i \right) \wedge (\ulen' > \eta)} \\
= \Prob{\Ppred = \mbox{ \rm true } \wedge \ulen > \eta}  \\
\le \Prob{\ulen > \eta} 
\end{gather*}
with respect to $\ang{\yvar_0, \Yvar_1\ldots,\Yvar_{2 \cdot \Fdata -1}}$,
thus Inequality~\eqref{repairer reduc eq} holds.

The rest of the proof bounds $\Prob{\ulen > \eta}$
for repairer or \lrepairer\ $\Rfnc$, which provides the bound on 
Inequality~\eqref{repairer reduc eq}.
It can be verified that
\begin{equation}
\label{lenf exp eq}
\Exp{\ulen_i} = \frac{\rlen_i - \sum_{\ell=1}^{i-1} \ulen_{\ell}}{\Ndata-i}
\end{equation}
with respect to $\ang{\yvar_0, \ldots\yvar_{i-1},\Yvar_i}$.
Let \[ \rho = \frac{(1-\epscore) \cdot \clen}{2\cdot \Fdata -1}, \]
\[ \tau_i = \sum_{\ell=1}^i \ell = \frac{i\cdot(i+1)}{2}. \]
If 
\begin{equation}
\label{lenf big eq}
\sum_{\ell=1}^{i-1} \ulen_\ell \ge \tau_{i-1} \cdot \rho
\end{equation}
with respect to $\ang{\yvar_0, \ldots,\yvar_{i-1}}$
then 
\begin{equation}
\label{lenf bound eq}
\Exp{\ulen_i} \le i \cdot \rho
\end{equation}
with respect to $\ang{\yvar_0, \ldots,\yvar_{i-1}, \Yvar_i}$.
This follows from Equation~\eqref{lenf exp eq},
Condition~\eqref{repairer less eq}, Inequality~\eqref{lenf big eq},
and because
\[ \frac{\Gamma_i -  \tau_{i-1} \cdot \rho}{\Ndata-i} = i \cdot \rho. \]
Define $\zvar_0=0$, and for $i = 1,\ldots,2 \cdot\Fdata-1$,
\begin{equation}
\label{zdata defn eq}
\Zvar_i  = \zvar_{i-1} + \ulen_i - i \cdot \rho  = \sum_{\ell=1}^i \ulen_{\ell} - \tau_i \cdot \rho
\end{equation}
with respect to $\ang{\yvar_0, \ldots,\yvar_{i-1}, \Yvar_i}$,
and define $\zvar_i$ similarly with respect to $\ang{\yvar_0, \ldots,\yvar_{i-1}, \yvar_i}$.
It can be verified that
\[ \tau_{2 \cdot \Fdata -1} \cdot \rho = \Fdata \cdot \clen - \epscore \cdot \Fdata \cdot \clen 
= \eta - (\epscore \cdot \Fdata-1) \cdot \clen, \]
thus 
\begin{equation}
\label{zdata eq}
\Prob{\Zvar_{2 \cdot \Fdata -1} > (\epscore \cdot \Fdata-1) \cdot \clen} = \Prob{\ulen > \eta}
\end{equation}
with respect to $\ang{\yvar_0, \Yvar_1,\ldots, \Yvar_{2 \cdot \Fdata - 1}}$.

It can be verified that 
\begin{equation*}
\abs{\zvar_i - \zvar_{i-1}} \le \clen
\end{equation*}
with respect to all $\ang{\yvar_0, \ldots,\yvar_{i-1}, \yvar_i}$.
Also, Equation~\eqref{zdata defn eq} and 
Inequalities~\eqref{lenf big eq} and \eqref{lenf bound eq} 
imply that if $\zvar_{i-1} \ge 0$ then
\begin{equation*}
\Exp{\Zvar_i} \le \zvar_{i-1}
\end{equation*}
with respect to $\ang{\yvar_0, \ldots,\yvar_{i-1}, \Yvar_i}$.
Thus, $\zvar_0, \Zvar_1,\ldots,\Zvar_{2\cdot \Fdata-1}$ with respect to
$\ang{\yvar_0, \Yvar_1,\ldots, \Yvar_{2 \cdot \Fdata -1}}$ satisfies the conditions
of \Supertheorem\ of Subsection~\ref{super sec} 
with $n=2\cdot F-1$, $c = \clen$, 
and $\alpha = (\epscore \cdot \Fdata - 1) \cdot \clen$.
Thus, from \Supertheorem\
and Equation~\eqref{zdata eq}, it can be verified that
\[ \Prob{\ulen > \eta} \le \delcore. \]
 with respect to
$\ang{\yvar_0, \Yvar_1,\ldots, \Yvar_{2 \cdot \Fdata -1}}$.
The lemma follows from Inequality~\eqref{repairer reduc eq}.
\qed
\end{proof}
With the settings in Section~\ref{practical setting sec} and $\betap = 0.1$,
$\delcore \le 3 \cdot 10^{-7}$ when $\epscore = 0.1$,  and
$\delcore \le 2 \cdot 10^{-39}$ when $\epscore = 0.2$.

\vspace{0.1in}
\begin{coretheorem}
\label{core theorem}
Fix $\betap < 1/2$. Fix $\epscore$ with $0 \le \epscore \le 1$,
and Equation~\eqref{delcore eq} defines $\delcore$.
For any repairer $\Rfnc$ and 
recoverer $\Afnc$, for any fixed $\tseq$,
with probability at least 
\[ 1-\left( \delcore + 2^{-\clen} \right), \]
with respect to $\Xvar$ and $(\tseq,\Yseq)$, 
at least one of the following two statements is true:
\begin{description}
\item{(1)}
There is an $i \in \{1,\ldots,M-1\}$ such that the average number of \bitval s read by the repairer
between $\tvarm_0$ and $\tvarm_i$ per each of the $i$
\nfail s is at least 
\begin{equation}
 (1-\epscore) \cdot \frac{(1-\betap) \cdot \clen}{2 \cdot \betap}. \label{core theorem eq}
\end{equation}
\item{(2)}
\Srcdata\ $\Xvar$ is unrecoverable by $\Afnc$ at time
$\tvarp_{M-1}$.
\end{description}
\end{coretheorem}
\begin{proof}
Apply \Corelemma\ with respect to all $\xvar$
and $(\tseq,\Yseq)$.  Apply \Compcorollary\ with
\begin{equation*}
\Xset = \left\{ \xvar: 
\left(\forall_{i=1}^{2 \cdot \Fdata -1}\rlen_{i} < 
\Gamma_i \right) \wedge 
(\ulen \le (\Fdata-1) \cdot \clen) \right\}
\end{equation*}
with respect to $\Xvar$ and any $(\tseq,\yseq)$.
Use 
\[\frac{\Gamma_i}{i} \ge \frac{\Gamma_{2 \cdot \Fdata-1}}{2 \cdot \Fdata-1} 
\ge (1-\epscore) \cdot \frac{(1-\betap) \cdot \clen}{2 \cdot \betap}\]
for any $i \in \{1,\ldots,M-1\}$, where Equation~\eqref{gamma eq} 
defines $\Gamma_i$.
\qed
\end{proof}

\subsection{Supermartingale bound}
\label{super sec}

We provide a probability bound used in the proof of 
\Corelemma\
that may be of independent interest.  Any improvement to this bound
provides an immediate improvement to \Corelemma.
We generalize previous notation. 
\vspace{0.1in}
\begin{supertheorem}
\label{super theorem}  
Let, $\zvar_0, \Zvar_1,\ldots,\Zvar_{n}$ 
be a random sequence of real-values
defined with respect to another random sequence
$\set{\yvar_0, \Yvar_1,\ldots, \Yvar_{n}}$, such that $z_0=0$ and the 
following conditions are satisfied for $i = 1,\ldots, n$.
\begin{itemize}
\item
$z_i$ is determined by $\set{\yvar_0, \yvar_1,\ldots, \yvar_{i}}$.
\item
$\abs{\zvar_i - \zvar_{i-1}} \le c$
with respect to all $\set{\yvar_0, \ldots,\yvar_{i-1}, \yvar_i}$.
\item
if $\zvar_{i-1} >  0$ then $\Exp{\Zvar_i} \le \zvar_{i-1}$
with respect to $\set{\yvar_0, \ldots,\yvar_{i-1}, \Yvar_i}$.
\end{itemize}
Then, for any $\alpha > 0$,
\begin{equation*}
\Prob{\Zvar_{n} > \alpha+c} \le n \cdot e^{\frac{-\alpha^2}{2 \cdot n \cdot c^2}}
\end{equation*}
\end{supertheorem}
\begin{proof}
For $\ell= 1,\ldots,n$, let predicate $\Ppred_\ell$ be
defined as follows on input 
$\set{\yvar_0,\ldots,\yvar_\ell,\ldots,\yvar_i}$, 
with $i \in \{ \ell,\ldots,n \}$. 
\begin{equation*}
\Ppred_\ell \mbox{ \rm is true }  \iff \zvar_{\ell-1} \le 0 \wedge \zvar_\ell > 0
\end{equation*} 
with respect to $\set{\yvar_0,\ldots,\yvar_\ell,\ldots,\yvar_i}$.  

For each $\ell= 1,\ldots,n$ and each $\set{\yvar_0,\ldots,\yvar_\ell}$ such
that $\Ppred_\ell$ is true, define a sequence as follows.
\begin{itemize}
\item
$\zvar^{\ell,\set{\yvar_0,\ldots,\yvar_\ell}}_\ell = \zvar_\ell$ 
with respect to $\set{\yvar_0,\ldots,\yvar_\ell}$.
\item
For $i = \ell+1,\ldots, n$, 
\begin{align}
\Zvar^{\ell,\set{\yvar_0,\ldots,\yvar_\ell}}_i & = \Zvar_{i} & 
{\rm if } \zvar^{\ell,\set{\yvar_0,\ldots,\yvar_\ell}}_{i-1} > 0 \label{super 1 eq}\\
\Zvar^{\ell,\set{\yvar_0,\ldots,\yvar_\ell}}_i & = 
\zvar^{\ell,\set{\yvar_0,\ldots,\yvar_\ell}}_{i-1} & 
{\rm if } \zvar^{\ell,\set{\yvar_0,\ldots,\yvar_\ell}}_{i-1} \le 0 \label{super 2 eq}
\end{align}
with respect to 
$\set{\yvar_0, \ldots,\yvar_\ell, \yvar_{\ell+1},\ldots, \yvar_{i-1}, \Yvar_i}$.
\end{itemize}
It can be verified that, for all 
$\set{\yvar_0, \ldots,\yvar_\ell, \yvar_{\ell+1},\ldots, \yvar_i}$,
\begin{equation}
\label{super bound eq}
\abs{\zvar^{\ell,\set{\yvar_0,\ldots,\yvar_\ell}}_i - 
\zvar^{\ell,\set{\yvar_0,\ldots,\yvar_\ell}}_{i-1}} \le c
\end{equation}
with respect to $\set{\yvar_0, \ldots,\yvar_\ell, \yvar_{\ell+1},\ldots, \yvar_i}$.

With respect to $\set{\yvar_0, \ldots,\yvar_\ell, \yvar_{\ell+1},\ldots, \yvar_{i-1},\Yvar_i}$:
Equations~\eqref{super 1 eq} and~\eqref{super 2 eq} imply 
that $\zvar^{\ell,\set{\yvar_0,\ldots,\yvar_\ell}}_{i-1} =  \zvar_{i-1}$
if $\zvar^{\ell,\set{\yvar_0,\ldots,\yvar_\ell}}_{i-1} > 0$, 
and since $\Exp{\Zvar_i} \le \zvar_{i-1}$ if $\zvar_{i-1} > 0$, it follows that
\[ \Exp{\Zvar^{\ell,\set{\yvar_0,\ldots,\yvar_\ell}}_i} \le 
\zvar^{\ell,\set{\yvar_0,\ldots,\yvar_\ell}}_{i-1} \]
if $\zvar^{\ell,\set{\yvar_0,\ldots,\yvar_\ell}}_{i-1} > 0$.
From Equation~\eqref{super 2 eq}, 
\[ \Exp{\Zvar^{\ell,\set{\yvar_0,\ldots,\yvar_\ell}}_i} = \zvar^{\ell,\set{\yvar_0,\ldots,\yvar_\ell}}_{i-1} \] 
if $\zvar^{\ell,\set{\yvar_0,\ldots,\yvar_\ell}}_{i-1} \le 0$.
Thus, 
\begin{equation}
\label{super exp eq}
\Exp{\Zvar^{\ell,\set{\yvar_0,\ldots,\yvar_\ell}}_i} \le  
\zvar^{\ell,\set{\yvar_0,\ldots,\yvar_\ell}}_{i-1}
\end{equation}
with respect to 
$\set{\yvar_0, \ldots,\yvar_\ell, \yvar_{\ell+1},\ldots, \yvar_{i-1},\Yvar_i}$.

From Equations~\eqref{super bound eq} and~\eqref{super exp eq}, 
for $\ell = 1,\ldots, n$, for all $\set{\yvar_0, \ldots, \yvar_{\ell}}$ 
such that $\Ppred_\ell$ is true,
\[ \zvar^{\ell,\set{\yvar_0,\ldots,\yvar_\ell}}_\ell, 
\Zvar^{\ell,\set{\yvar_0,\ldots,\yvar_\ell}}_{\ell+1},\ldots,
\Zvar^{\ell,\set{\yvar_0,\ldots,\yvar_\ell}}_{n} \] 
with respect to
$\set{\yvar_0, \ldots, \yvar_\ell, \Yvar_{\ell+1}, \ldots, \Yvar_{n}}$
is a supermartingale.
Thus, from the Azuma's inequality, 
\begin{equation}
\label{azuma eq}
\Prob{\Zvar^{\ell,\set{\yvar_0,\ldots,\yvar_\ell}}_{n} - \zvar^{\ell,\set{\yvar_0,\ldots,\yvar_\ell}}_\ell > \alpha} \le e^{\frac{-\alpha^2}{2 \cdot (n-\ell) \cdot c^2}}
\end{equation}
with respect to $\set{\yvar_0, \ldots,\yvar_\ell, \Yvar_{\ell+1},\ldots, \Yvar_{n}}$.
It can be verified that $\zvar^{\ell,\set{\yvar_0,\ldots,\yvar_\ell}}_\ell \le c$ if $\Ppred_\ell$ is true for 
$\set{\yvar_0,\ldots,\yvar_\ell}$, thus
\begin{gather}
\label{martin corr eq}
\Prob{\Zvar^{\ell,\set{\yvar_0,\ldots,\yvar_\ell}}_{n} > \alpha + c} \\
\le \Prob{\Zvar^{\ell,\set{\yvar_0,\ldots,\yvar_\ell}}_{n} - \zvar^{\ell,\set{\yvar_0,\ldots,\yvar_\ell}}_\ell > \alpha} \nonumber
\end{gather}
with respect to $\set{\yvar_0, \ldots,\yvar_\ell, \Yvar_{\ell+1},\ldots, \Yvar_{n}}$.

It can be verified that, for any 
$\set{\yvar_0, \ldots, \yvar_{n}}$,
\begin{gather}
\label{super corr eq}
\zvar_n > \alpha + c \iff \\
\exists_{\ell = 1}^{n} \mbox{ \rm s.t. }
\Ppred_\ell \mbox{ \rm is true } \wedge 
\zvar^{\ell,\set{\yvar_0,\ldots,\yvar_\ell}} > \alpha + c \nonumber
\end{gather}
with respect to $\set{\yvar_0, \ldots, \yvar_{n}}$.
From Equation~\eqref{super corr eq} it follows that
\begin{gather}
\label{giant eq}
\Prob{\Zvar_n > \alpha + c} \le \\ 
\sum_{\ell=1}^n \Prob{\Ppred_\ell \mbox{ \rm is true }  \wedge 
\Zvar^{\ell,\set{\yvar_0, \Yvar_1,\ldots, \Yvar_{\ell}}}_n > \alpha+c} \nonumber \\
\end{gather}
with respect to $\set{\yvar_0,\Yvar_1, \ldots, \Yvar_{n}}$.
From Inequalites~\eqref{giant eq}, \eqref{martin corr eq}, \eqref{azuma eq},
it follows that 
\begin{gather}
\Prob{\Zvar_n > \alpha + c} \le  \sum_{\ell=1}^n 
e^{\frac{-\alpha^2}{2 \cdot (n-\ell) \cdot c^2}} \le 
n \cdot e^{\frac{-\alpha^2}{2 \cdot n \cdot c^2}},
\end{gather}
with respect to $\set{\yvar_0,\Yvar_1, \ldots, \Yvar_{n}}$.
\qed
\end{proof}

\section{Main lower bounds}
\label{main sec}

All the \nfail s within a phase are distinct in
Section~\ref{core lower sec}, and thus the analysis
does not apply to \nfseq s where the \nfail s are
independent.  This section extends the results to
random and independent \nfseq s.  

\Unitheorem\ presented below
shows lower bounds with respect to the uniform
\idseq\ for any fixed \timeseq\ $\tseq$.
The uniform \idseq\ for a phase with $M$ distinct
\nfail s can be generated from
the distributions $\Gseq$ and $\Yseq$ described below. 

Let $\Gvar_i$ be an independent geometric 
random variable with success probability $\frac{\Ndata-i}{\Ndata}$,
and thus $\Exp{\Gvar_i} = \frac{\Ndata}{\Ndata-i}$, and let 
\[ \Gseq = \{\Gvar_1,\ldots,\Gvar_{M-1} \}. \] 
As before, let 
\[ \Yseq = \ang{\yvar_0,\Yvar_1,\ldots,\Yvar_{M-1}} \]
be a random distinct \idseq\ for the phase.

The uniform \idseq\ $\Useq$ for the phase 
can be generated as follows from $\Gseq$ and $\Yseq$.
Let $\gsum_0 = 0$.
For $i = 1,\ldots, M-1$, let 
\[ \gsum_i = \sum_{j=1}^i \Gvar_j. \]
For $i \in \{ 1,\ldots,M-1 \}$, let
\[ \Uvar_{\gsum_i} = \Yvar_i, \]
and for $j = \gsum_{i-1}+1,\ldots,\gsum_i-1$, let
\[ \Uvar_j \inu \{\yvar_0,\Yvar_1,\ldots,\Yvar_{i-1} \}. \]
Then,
\[ \Useq = \{ \yvar_0,\Uvar_1,\Uvar_2,\ldots,\Uvar_{\gsum_{M-1}} \} \] 
is a random uniform \idseq.

Of interest in the proof of \Unitheorem\ is 
\begin{equation}
\label{exp geosum eq}
\Exp{\gsum_{i}}
= \sum_{j=1}^{i} \frac{\Ndata}{\Ndata-j},
\end{equation}
since this is the expected number of \nfail s in a phase until
there are $i$ distinct \nfail s beyond the initial \nfail.
For $0 \le \zeta < 1$, define
\begin{equation}
\label{lnifunction eq}
\lnifunction(\zeta) = \ln\left(\frac{1}{1-\zeta}\right).
\end{equation}
Note that
\begin{equation}
\label{lnifunction ineq}
\sum_{j=0}^{\zeta \cdot \Ndata-1} \frac{1}{\Ndata-j} < \lnifunction(\zeta \cdot \Ndata) < \sum_{j=1}^{\zeta \cdot \Ndata} \frac{1}{\Ndata-j}.
\end{equation}
Let 
\begin{equation}
\label{fpdata eq}
\Fpdata = 2 \cdot\Fdata\cdot \frac{\lnifunction(2 \cdot \betap)}{2 \cdot \betap} 
= \lnifunction(2 \cdot \betap) \cdot \Ndata.
\end{equation}
It follows from Equations~\eqref{exp geosum eq}
 and~\eqref{lnifunction ineq} that
\[ \Exp{\gsum_{2 \cdot \Fdata-1}} \le \Fpdata. \]
Because $\lnifunction(\zeta) \rightarrow \zeta$ as $\zeta \rightarrow 0$,
$\Fpdata \rightarrow 2 \cdot \Fdata$ as $\betap \rightarrow 0$.

One can view $\Gseq$ as being part of the \timeseq\ for a phase as follows.
For $i = 1,\ldots, 2 \cdot \Fdata -1$, let $\Thvar_i = \tvar_{\gsum_i}$
be the time of the $i^{\rm th}$ distinct \nfail\ beyond the initial \nfail.
 This defines a \timeseq\
 \[ \Tpseq = \{\tvar_0,\Thvar_1,\ldots,\Thvar_{2 \cdot \Fdata -1}\}, \]
and the corresponding distinct \nfail s at these times are
\[ \Yseq = \ang{\yvar_0,\Yvar_1,\ldots,\Yvar_{M-1}}, \]
where $\Tpseq$ is defined by $(\tseq,\Gseq)$ and is independent of $\Yseq$.

A phase proceeds as follows with respect to \srcdata\ $\xvar$. 
For each $i = 1,\ldots, 2 \cdot \Fdata-1$, 
$\Gvar_i$ is chosen independently and
$\Rfnc$ is executed up till time $\Thvarm_i$.  If 
$\rlen_i \ge \Gamma_i$ with respect to $\xvar$ and 
$\{\tvar_0,\Thvar_1,\ldots,\Thvar_{i}\}$ and 
$\{\yvar_0,\Yvar_1,\ldots,\Yvar_{i-1}\}$
then the phase ends at time $\Thvar_i$.
If the phase doesn't end in the above process then
$\rlen_i < \Gamma_i$ for  $i =1,\ldots,  2 \cdot \Fpdata-1$,
and the phase ends at time $\Thvar_{2\cdot F-1}$.  

The condition $\rlen_i \ge \Gamma_i$ doesn't provide a guarantee
that the amount of data read by $\Rfnc$ up till time $\Thvar_i$ is
$\frac{\Gamma_i}{i}$ per \nfail, since the number of \nfail s 
up till $\Thvar_i$ is $\gsum_i$, which is a random variable that can be highly variable.  
To be able to prove that $\Rfnc$ reads a lot of data per \nfail, we stitch phases 
together into a sequence of phases, and argue that $\Rfnc$ must read 
a lot of data per \nfail\ over a sequence of phases that
covers a large enough number of distinct \nfail s.
\Geolemma\ below provides the technical details.

For $-1 < \zeta < 1$, define
\begin{equation}
\label{lndfunction eq}
\lndfunction(\zeta) = \zeta - \ln(1+\zeta).
\end{equation}
Note that $\lndfunction(\zeta) \rightarrow \frac{\zeta^2}{2}$ as $\zeta \rightarrow 0$.

\vspace{0.1in}
\begin{geolemma}
\label{geo lemma}
Fix $\betap < 1/2$. Fix $\epsgeo > 0$ and let 
\begin{equation*}
\delgeo =  2 \cdot \Fdata \cdot  
\frac{e^{-2 \cdot \betap \cdot (1-2\cdot \beta') \cdot \Ndata
\cdot \lndfunction(\epsgeo)}}{1+\epsgeo}.
\end{equation*} 
For any repairer $\Rfnc$ and 
recoverer $\Afnc$, for any $\xvar$ and $\tseq$, 
with probability at least $1-\delgeo$
with respect to $\xvar$ and $(\tseq,\Useq)$, 
there is an $m \le (1+\epsgeo) \cdot 2 \cdot  \Fpdata$
such that a phase ends at $\tvar_m$
and the number of distinct \nfail s in the sequence of phases between
$\tvarp_0$ and $\tvarp_m$  is at least 
\begin{equation}
\frac{2 \cdot \betap}
{(1+\epsgeo) \cdot \lnifunction(2 \cdot \betap)} \cdot m. \label{geo theorem eq}
\end{equation}
\end{geolemma}
\begin{proof}
Fix a $\ell$ with $2 \cdot \Fdata \le \ell \le 4 \cdot \Fdata$.
Stitch together phases into a sequence of phases starting at time $\tvar_0$
until exactly $\ell$ distinct \nfail s have occurred
(this may occur in the middle of a phase). Let $\pvar$
be the number of phases including the last possibly partially completed phase.
For $j=1,\ldots,\pvar$, let $\dvar_j$ be the number of distinct \nfail s in phase $j$,
let
\[ \Gseq_j = \Gvar_{j,1},\ldots,\Gvar_{j,\dvar_j} \]
be the geometric random variables in phase $j$,
where $\Gvar_{j,i}$ has success probability $\frac{\Ndata-i}{\Ndata}$,
and let 
\[ \Gseq = \Gseq_1, \Gseq_2, \Gseq_\pvar  \]
be the geometric random variables in the sequence of phases.
Note that $\sum_{j=1}^\pvar \dvar_j = \ell$, and
\[ \sum_{j=1}^{\pvar} \sum_{i=1}^{\dvar_j} \Gvar_{j,i} \]
is the number of \nfail s in the sequence of phases.

The indices of the random variables in $\Gseq$ depend on the
history of the process, i.e., if the current index is $i$ in phase $j$
then the next index is $i+1$ if the phase doesn't complete, 
and the next index is $1$ if the phase does complete and the next
phase $j+1$ starts, with the restriction that the index never exceeds 
$2 \cdot \Fdata-1$.  However, $\Gvar_{j,i}$ is chosen independently 
of all previous history once index $i$ in phase $j$ is determined.
Thus, $\Gseq$ is a sequence of independent random variables,
but which random variables are in $\Gseq$ depends on the process.

Let $\phvar = \lceil \frac{\ell}{2 \cdot \Fdata-1}\rceil,$
for $j = 1,\ldots,\phvar-1$ let $\dhvar_j = 2 \cdot \Fdata-1$,
let $\dhvar_{\phvar} = \ell- (2 \cdot \Fdata-1) \cdot (\phvar-1)$, define
a sequence of $\ell$ independent geometric random variables
\[ \Gpseq = \Gpseq_1, \Gpseq_2, \Gpseq_{\phvar},  \]
where 
\[ \Gpseq_j = \Gpvar_{j,1},\ldots,\Gpvar_{j,\dhvar_j}, \]
and $\Gpvar_{j,i}$ 
has success probability $\frac{\Ndata-i}{\Ndata}$.

As the sequence $\Gseq = \Gseq_1, \Gseq_2, \Gseq_\pvar$ of $\ell$ geometric
random variables defined by the process above is being generated,
after index $i$ of phase $j$ has been determined, $\Gvar_{j,i}$ 
can be matched with an unmatched $\Gpvar_{j',i'}$ of $\Gpseq$,
where 
\[ i' = \argmin_{i'} \set{ i' \ge i: \exists_{j'} \mbox{ s.t. } 
 \Gpvar_{j',i'} \mbox{ is unmatched}}. \] 
That a match can always be made can be shown based on the observation
that, for all $i$, 
\[ \abs{\set{\Gvar_{j,i'} \in \Gseq \mbox{ s.t. } i' \ge i}} 
\le \abs{\set{\Gpvar_{j,i'} \in \Gpseq \mbox{ s.t. } i' \ge i}}. \]
This holds independent of which random
variables are added to $\Gseq$ by the process. 
Only a prefix of $\Gseq$ is known at the time of each 
match, but all of $\Gpseq$ is known a priori. 

Random variable $\Gvar_{j,i}$ can be defined as
\[ \Gvar_{j,i} = \argmin_n \left\{ \Bvar_n \le \frac{\Ndata-i}{\Ndata} \right\},\]
where $\{\Bvar_{1},\Bvar_{2},\ldots \}$ is an auxiliary set 
of independent random variables uniformly distributed in $[0,1]$.
Each matched pair $\Gvar_{j,i}$ and $\Gpvar_{j',i'}$, 
can be defined by the same auxiliary set,
and from $i' \ge i$ it follows that $\Gpvar_{j',i'} \ge \Gvar_{j,i}$ 
for all possible values of the auxiliary set.
Thus,
\begin{equation}
\label{dom eq}
\Probi{\Gseq}{\sum_{j=1}^\pvar \sum_{i=1}^{\dvar_j} \Gvar_{j,i}\ge \eta} \le 
\Probi{\Gpseq}{\sum_{j=1}^{\phvar} \sum_{i=1}^{\dhvar_j} \Gpvar_{j,i} \ge \eta},
\end{equation}
 for any $\xvar$ and $(\tseq,\yseq)$.  With
 \begin{equation}
 \label{ellp eq}
 \ell' = \ell \cdot \frac{\lnifunction(2 \cdot \betap)}{2 \cdot \betap},
 \end{equation}
it can be verified from Equation~\eqref{lnifunction ineq} that 
\begin{equation}
\label{geosum eq}
\Exp{\sum_{j=1}^{\phvar} \sum_{i=1}^{\dhvar_j} \Gpvar_{j,i}} \le \ell'.
\end{equation}
Let 
\begin{equation*}
\delgeo' = \frac{e^{-2 \cdot \betap \cdot (1-2\cdot \beta') \cdot \Ndata
\cdot \lndfunction(\epsgeo)}}{1+\epsgeo}.
\end{equation*}
From Equation~\eqref{geosum eq}, using 
\[ \ell' \ge \Fpdata \ge 2 \cdot \Fdata = 2 \cdot \betap \cdot \Ndata, \] 
and from Theorem 2.1 of~\cite{Janson17},
it follows that
\begin{equation}
\label{geo eq}
\Prob{\sum_{j=1}^{\phvar} \sum_{i=1}^{\dhvar_j} \Gpvar_{j,i}  \ge (1+\epsgeo) \cdot \ell'}
\le \delgeo'.
\end{equation}

Consider executing the process until there are $2\cdot \Fdata$ 
distinct \nfail s, and then continuing the process until the current phase completes
when there are some number $\ell$ of distinct \nfail s total, where $\ell$
is no longer fixed but instead determined by the process, and let $m$ be
the total number \nfail s determined by the process.  
Since no phase has more than $2\cdot \Fdata-1$ distinct \nfail s,
it follows that $\ell \le 4 \cdot \Fdata$ distinct \nfail s.  
Since there are at most $2 \cdot \Fdata$ possible values for $\ell$, 
and from a union bound and using Equations~\eqref{geo eq},
\eqref{ellp eq}, and~\eqref{dom eq}, 
\[ m \ge (1+\epsgeo) \cdot 
\frac{\lnifunction(2 \cdot \betap)}{2 \cdot \betap} \cdot \ell \]
with probability at most $\delgeo = 2 \cdot \Fdata \cdot \delgeo'$.
Thus, with probability at least $1-\delgeo$,
\[ \ell \ge \frac{2 \cdot \betap} {(1+\epsgeo) 
\cdot \lnifunction(2 \cdot \betap)} \cdot m. \]
This also shows that, with probability at least $1-\delgeo$,
\[ m \le 2 \cdot  (1+\epsgeo) \cdot \Fpdata, \]
since 
$\ell \le 4 \cdot \Fdata = 2 \cdot \Fpdata \cdot \frac{2 \cdot \betap}{\lnifunction(2 \cdot \betap)}$ from Equation~\eqref{fpdata eq}.  
\qed
\end{proof}

\subsection{Uniform failures lower bound}
\label{uniform lower bound sec}

\vspace{0.1in}
\begin{unitheorem}
\label{uniform theorem}
Fix $\betap < 1/2$. Let $\epscore$ and $\delcore$ be as defined
in \Corelemma, and let $\epsgeo$ and $\delgeo$ be as defined in
\Geolemma, and let 
\[ \deluni = \delgeo + \Fdata \cdot (\delcore + 2^{-\clen}). \]
For any repairer $\Rfnc$ and 
recoverer $\Afnc$, for any fixed $\tseq$,
with probability at least  $1-\deluni$
with respect to $\Xvar$ and $(\tseq,\Useq)$, 
at least one of the following two statements is true:
\begin{description}
\item{(1)}
There is an $m \le (1+\epsgeo) \cdot 2 \cdot \Fpdata$ such that the average number of \bitval s read by the repairer between $\tvarm_0$ and $\tvarm_m$ per each of the $m$
\nfail s is at least
\begin{equation}
 \frac{(1-\epscore)}{(1+\epsgeo)} \cdot
 \frac{(1-\betap) \cdot \clen}{\lnifunction(2 \cdot \betap)}. \label{uniform theorem eq}
\end{equation}
\item{(2)}
\Srcdata\ $\Xvar$ is unrecoverable by $\Afnc$ at time
$\tvarp_{m}$.
\end{description}
\end{unitheorem}
\begin{proof}
From \Geolemma, there is a sequence of phases that ends with 
$m \le (1 + \epsgeo) \cdot  2 \cdot \Fpdata$ \nfail s where 
the number of distinct \nfail s is at least  
Equation~\eqref{geo theorem eq} with probability at least $1-\delgeo$.
From \Coretheorem\ with respect to all $\xvar$
and $(\tseq,\Gseq,\Yseq)$ and using a union bound over at most $\Fdata$
phases in the sequence of phases, the average number of \bitval s read
by $\Rfnc$ between $\tvarm_0$ and $\tvarm_m$ per each
distinct \nfail\ is at least Equation~\eqref{core theorem eq} with probability
at least $1-\Fdata \cdot (\delcore + 2^{-\clen})$. 
Thus, overall the two statements hold with probability at
least $1-\deluni$.
\qed
\end{proof}

Since the end of one sequence of phases can be the beginning of the next sequence
of phases, it follows that the average number of \bitval s read by the repairer 
per \nfail\ must satisfy Equation~\eqref{uniform theorem eq} 
over the entire lifetime of the system for which the \srcdata\ is recoverable.

Equation~\eqref{uniform theorem eq} holds independent of the \timeseq.
Thus, if there are a lot of \nfail s over a period of time then the
\rrepairrate\ over this period of time must necessarily be high, whereas
if there are fewer \nfail s over a period of time then the \rrepairrate\ over
this period of time can be lower.   Automatic adjustments of the \rrepairrate\
as the \nfail\ rate fluctuates is one of the key contributions of the algorithms
described in \cite{Luby16}, which shows that there are algorithms that can
match the lower bounds of \Unitheorem, even for a fluctuating \timeseq.

\subsection{Poisson failures lower bound}
\label{poisson lower bound sec}

\Unitheorem\ expresses lower bounds in terms of the average number of \bitval s
read per \nfail.  \Poissonthm, presented in this section, instead expresses
the lower bounds in terms of a \rrepairrate.  The primary additional
technical component needed to prove \Poissonthm\ is a concentration
in probability result: the number of \nfail s for a \Ptdist\ with rate $\lambda$
over a suitably long period of time is relatively close to the 
expected number of \nfail s with high probability.

The \Ptdist\ with rate $\lambda$ can be generated as follows. 
For $i \ge 1$, let $\Qvar_i$ be an independent exponential random variable 
with rate $\lambda \cdot \Ndata$,
and let 
\[ \Qseq = \{ \Qvar_1,\ldots,\Qvar_{M-1} \}. \]
For $i \ge 1$, let
\[ \Tvar_i = \tvar_0 + \sum_{j=1}^i \Qvar_j, \]
and let 
\[ \Tseq = \{\tvar_0, \Tvar_1,\ldots,\Tvar_{M-1} \}. \]
For $i \ge 1$, let $\Uvar_i$ be an independent random variable that is
uniformly distributed in $\{0,\ldots,\Ndata-1\},$
and let 
\[ \Useq = \{\yvar_0, \Uvar_1,\ldots,\Uvar_{M-1} \}. \]
Then, $(\Tseq,\Useq)$ is a random \nfseq\ 
with respect to the \Ptdist\ with rate $\lambda$.

Capacity is erased from the system at a rate 
\[ \erate = \lambda \cdot \Ndata \cdot \clen \]
with respect to the \Ptdist\ with rate $\lambda$. 

\vspace{0.1in}
\begin{Poissontheorem}
\label{Poisson theorem}
Fix $\betap < 1/2$.  Let $\epscore$ be as defined
in \Corelemma, $\epsgeo$ be as defined in \Geolemma,
and $\deluni$ be as defined in \Unitheorem. Let $\epspoi > 0$, let 
\begin{equation*}
\delpoi =  \deluni + (1+\epsgeo) \cdot 2 \cdot \Fpdata 
\cdot  \frac{e^{- 2 \cdot \Fdata \cdot \lndfunction(\epspoi) }}{1+\epspoi}.
\end{equation*}
Let the \nfseq\ be a \Ptdist\ with rate $\lambda$, and let
\begin{equation*}
\Delta = (1+\epsgeo) \cdot (1+\epspoi) \cdot \frac{2 \cdot \lnifunction(2 \cdot \betap)}{\lambda}.
\end{equation*}
For any repairer $\Rfnc$ and 
recoverer $\Afnc$, for any starting time $\tvar_0$,
with probability at least  $1-\delpoi$, 
at least one of the following two statements is true:
\begin{description}
\item{(1)}
There is a $\tvar \le \tvar_0 + \Delta$ 
such that the average rate $\rrate$ the repairer reads \bitval s between 
$\tvar_0$ and $\tvar$ satisfies
\begin{equation}
\rrate \ge \frac{(1-\epscore)}{(1+\epsgeo) \cdot (1+\epspoi)} \cdot
 \frac{(1-\betap)}{\lnifunction(2 \cdot \betap)} 
 \cdot \erate. \label{Poisson theorem eq}
\end{equation}
\item{(2)}
\Srcdata\ $\Xvar$ is unrecoverable by $\Afnc$ at time $\tvar_0 + \Delta$.
\end{description}
\end{Poissontheorem}
\begin{proof}
From \Unitheorem, there is a sequence of phases that ends with 
$m \le (1 + \epsgeo) \cdot 2 \cdot \Fpdata$ \nfail s where 
the number of distinct \nfail s is provided by 
Equation~\eqref{uniform theorem eq} with probability at least $1-\deluni$.
Since there are at least $2 \cdot \Fdata$ distinct \nfail s in the process, 
$m \ge 2 \cdot\Fdata$.

For each $m'$ between $2 \cdot \Fdata$ and  $(1+\epsgeo) \cdot 2 \cdot \Fpdata$,
when
\[\delpoi' (m') = \frac{e^{-m' \cdot \lndfunction(\epspoi)}}{1+\epspoi}, \]
it follows from Theorem~5.1 of \cite{Janson17} that
\begin{equation*}
\Prob{\sum_{i=1}^{m'} \Qvar_i \ge (1+\epspoi) \cdot \frac{m'}{\lambda \cdot N}}
\le \delpoi'(m').
\end{equation*}
Using a union bound, it follows that with probability at least 
$1-(1+\epsgeo) \cdot 2 \cdot \Fpdata\cdot \delpoi'(2 \cdot \Fdata)$,
\[
\sum_{i=1}^m \Qvar_i < (1+\epspoi) \cdot \frac{m}{\lambda \cdot N}.
\]
Thus, the time $\tvar = \tvar_0 + \thvar$ when there 
are $m$ \nfail s in the process satisfies
\begin{equation}
\label{time eq}
\thvar \le \frac{(1+\epspoi) \cdot m}{\lambda \cdot \Ndata}
\end{equation}
with probability at least 
$1-(1+\epsgeo) \cdot 2 \cdot \Fpdata\cdot \delpoi'(2 \cdot \Fdata)$.

From \Unitheorem, and combining Equations~\eqref{uniform theorem eq}
and~\eqref{time eq}, it follows that with probability at least $1-\delpoi$
the rate at which the repairer reads data between $\tvar_0$ and $\tvar$
is at least that in Equation~\eqref{Poisson theorem eq} or else the \srcdata\
is unrecoverable at time $\tvar = \tvar_0 + \thvar$, and thus unrecoverable at time
$\tvar_0 + \Delta \ge  \tvar_0 + \thvar$.
\qed
\end{proof}

Note that $\delpoi$ shrinks exponentially fast as $\Ndata$ grows.
Also, $\epscore$, $\epsgeo$, and $\epspoi$ can all approach zero as $\Ndata$ grows.
Thus, as $\Ndata$ grows, the lower bound on $\rrate$ of \Poissonthm\
can be expressed asymptotically as
\begin{equation}
\label{read rate eq}
\frac{\rrate}{\erate} \ge \frac{1-\betap}{\lnifunction(2 \cdot \betap)}.
\end{equation}
Since the end of one interval can be the beginning of the next interval, 
it follows that $\rrate$ must also satisfy Equation~\eqref{read rate eq} 
over the entire lifetime of the system.
Furthermore, as $\beta \rightarrow 0$, the lower bound approaches
\begin{equation}
\label{asymp read rate eq}
\frac{\rrate}{\erate} \ge \frac{1}{2 \cdot \beta}.
\end{equation}

\subsection{Distributed storage \srcdata\ capacity}
\label{information sec}

Our distributed storage model and results are inspired by Shannon's communication model~\cite{CShannon48}.  We define
the distributed storage \srcdata\ capacity to be
the amount of \srcdata\ that can be reliably stored 
for long periods of time by the system.
Equation~\eqref{fundamental eq} of the abstract is the asymptotic
distributed storage \srcdata\ capacity
as $\Ndata$ and $\rrate$ grow.
Equation~\eqref{asymp read rate eq} shows that the 
distributed storage \srcdata\ capacity cannot
asymptotically be larger than Equation~\eqref{fundamental eq},
and Equation~\eqref{alglimit eq} just after \ALiqpoissonthm\ shows 
that distributed storage \srcdata\ capacity asymptotically approaching Equation~\eqref{fundamental eq} can be achieved.

\section{\Liqrepairer}
\label{liqrepairer sec}

We first describe a simplified description and analysis
of the \liqrepairer\ described in~\cite{Luby16}, 
where the tradeoff between $\beta$ 
and the repairer read rate
is around a factor of two worse than optimal.  This provides
some intuition for the \advliqrepairer\ described in Section~\ref{advliqrepairer sec}.

The repairers we describe operate as follows.
The \srcdata\ is partitioned into objects, and an 
erasure code is applied separately to each object.
Descriptions of erasure codes are provided in~\cite{RFC5510}, \cite{RFC6330}.
The parameters of an erasure code are $(\ndata,\kdata,\rdata)$.
An object of size $\olen$ is partitioned into $\kdata$ source fragments,
each of size $\flen = \frac{\olen}{\kdata}$, and the erasure
encoder is used to generate up to $\rdata$ repair fragments, 
each of size $\flen$, from the $\kdata$ source fragments.
An {\em Encoding Fragment ID}, or EFI,  
is used to uniquely identify each fragment of an object,
where typically EFIs $0, \ldots, \kdata-1$ are used to 
identify the source fragments, and EFIs $\kdata,\ldots, \ndata-1$
are used to identify the repair fragments.
The erasure decoder is used to recover the object from a subset
of the source and repair fragments identified by their EFIs.
We assume any $\kdata$ fragments can be used to decode the object. 
This holds for Reed-Solomon~\cite{RFC5510}, 
and essentially holds for RaptorQ~\cite{RFC6330}.

\subsection{\Liqrepairer\ with periodic failures}
\label{liq periodic sec}

\vspace{.1in}
\begin{Liqperiodictheorem}
\label{Liqperiodic theorem}
Fix $\beta > 0$ and let $(1-\beta) \cdot \Ndata \cdot \clen$
be the size of the \srcdata.
A \liqrepairer\ with respect to a periodic timing \nfseq\
ensures the \srcdata\ is always recoverable,
with $\frac{1-\beta}{\beta} \cdot \clen$
data read and at most $\clen$ data written per \nfail.
\end{Liqperiodictheorem}

\begin{proof}
A \liqrepairer\ operates as follows. 
Let $\kdata = (1-\beta) \cdot \Ndata$,
and $\rdata = \beta \cdot \Ndata$. 
\Srcdata\ $\xvar$ of $\xlen$ \bitval s 
is partitioned into $\rdata$ equal-size objects 
$\xvar_0,\ldots,\xvar_{\rdata-1}.$
For $j= 0,\ldots, \rdata-1$, the storer uses an 
$(\Ndata,\kdata,\rdata)$-erasure code to generate fragments 
$i= 0,\ldots, \kdata +j$ from 
$\xvar_j$ and stores fragment $i$ for $\xvar_j$ at node $i$.
Thus, node $i$ stores all fragments with EFI $i$ for all objects.

The fragment size is $\flen = \frac{\clen}{\rdata}$, 
since each node has capacity to store a fragment for each of the $\rdata$ objects.
The \storeoverhead\ is $\beta$, since 
$\xlen = \rdata \cdot \kdata \cdot \flen = (1-\beta) \cdot \Ndata \cdot \clen$.

The invariant established by the storer and maintained
by the repairer is that, just before a \nfail, object
$\xvar_j$ has at least $\kdata + j+1$ stored fragments,
for $j = 0,\ldots, \rdata-1$. 

A \nfail\ decreases the number of stored fragments 
for an object by at most one.  Since the invariant ensures
at least $\kdata+1$ stored fragments 
for each object just before a \nfail, there are at least
$\kdata$ stored fragments for each object just after a \nfail,
and thus the \srcdata\ is recoverable at all times.

The repairer executes the following repair step
between \nfail s to reestablish the invariant.
The repairer reads $\kdata$ stored fragments 
for the object $\xvar_0$, uses the erasure decoder
to recover $\xvar_0$, uses the erasure encoder 
to generate the missing up to $\rdata$ fragments 
from $\xvar_0$, and stores these fragments at the nodes.  
Thus, the amount of data read per repair step is 
\begin{equation}
\label{rlen eq}
\rlen = \kdata \cdot \flen = 
\frac{1-\beta}{\beta} \cdot \clen,
\end{equation}
the the amount of data written per repair step is at most
\[ \wlen = \rdata \cdot \flen = \clen, \]
and repairer network traffic per step is at most 
$\tlen = \frac{\clen}{\beta}$.

After a repair step, $\xvar_0$ has $\ndata$ stored fragments.
The objects are logically reordered at this point,
i.e., for $j = 1,\ldots, \rdata-1$, the old $\xvar_j$ becomes the new $\xvar_{j-1}$, and
the old $\xvar_0$ becomes the new $\xvar_{\rdata-1}$. 
This reestablishes the invariant.
\qed
\end{proof}

As described in \cite{Luby16}, the \liqrepairer\ uses a data organization 
that allows encoding and decoding each object as a stream using as little as
$\Ndata \cdot \log(\Ndata)$ \bitval s of memory, even though an object
as described in the proof of \Liqperiodicthm\ is larger 
than the capacity of a node.

\subsection{\Liqrepairer\ with Poisson failures}
\label{liq poisson sec}

\vspace{.1in}
\begin{Liqpoissontheorem}
\label{Liqpoisson theorem}
Fix $\beta > 0$ and let $(1-\beta) \cdot \Ndata \cdot \clen$
be the size of the \srcdata.
For  $\epsliq > 0$, let
\begin{equation*}
\delliq = e^{-\frac{\epsliq^2 \cdot \left(1-\frac{\epsliq}{2}\right)
\cdot \beta \cdot \Ndata}{4}}.
\end{equation*}
Let the \nfseq\ be a \Ptdist\ with rate $\lambda$.
With probability at least $1-\Mdata^2\cdot \delliq$, a
\liqrepairer\ can maintain recoverability of 
$\xvar$ for $\Mdata$ \nfail s 
when the repairer peak read rate is
\begin{equation}
\label{liq upper eq}
\rrate \le \frac{1-\beta}{(1-\epsliq) \cdot \beta} 
\cdot \lambda \cdot \Ndata\cdot \clen.
\end{equation}
\end{Liqpoissontheorem}

\begin{proof}
Let $\epsliq' = \epsliq/2$.  
Set $\bdata = \epsliq' \cdot \rdata+1$, 
$\rpdata = (1-\epsliq') \cdot \rdata$
and $\betap = (1-\epsliq') \cdot \beta$.   Thus, $\rpdata + \bdata = \rdata + 1$.
A repair step is the same as in the proof of \Liqperiodicthm,
except that there are $\rpdata$ objects instead of $\rdata$,
and the fragment size is $\flen = \frac{\clen}{\rpdata}$, 
and thus the repairer read data per step is 
$\rlen = \frac{1-\beta}{\betap} \cdot \clen,$
the repairer write data per step is at most
$\wlen = \frac{\beta}{\betap} \cdot \clen,$
and repairer network traffic per step 
is at most $\tlen = \frac{\clen}{\betap}$.

Let $\bdata(\tvar)$ be a counter that is updated
as follows, where initially $\bdata(0) = \bdata$. 
\begin{itemize}
\item 
$\bdata(\tvarp) = \bdata(\tvarm)-1$ if there is a \nfail\ at $\tvar$. 
\item
$\bdata(\tvarp) = \min\{\bdata(\tvarm)+1,\bdata\}$ 
if a repair step ends at $\tvar$.
\end{itemize}

The invariant established by the storer and maintained
by the repairer is that object $\xvar_j$ has at least 
$\kdata +\bdata(\tvar)+ j$ stored fragments 
at time $\tvar$ for $j = 0,\ldots, \rpdata-1$.  
Thus, $\bdata(\thvar) \ge 0$ for $\thvar \le \tvar$ 
ensures the \srcdata\ is recoverable at time $\tvar$.

The invariant is reestablished after a \nfail\ at time $\tvar$, 
since each object loses at most one fragment and thus
each object has at most one less stored fragment due to the \nfail.

No repair step is executing at time $\tvar$ if $\bdata(\tvar) = \bdata$.  
If there is a \nfail\ at time $\tvar$ when $\bdata(\tvarm) = \bdata$, 
then $\bdata(\tvarp) = \bdata-1$ and a repair step is initiated.
Repair steps are executed sequentially, i.e. at most one repair step
is executing at time.  As soon as a repair step completes,
a next repair step can initiate.

The following shows that the invariant is reestablished 
for a repair step that starts at $\tvar$ 
when $\bdata(\tvarp) \le \bdata-1$ and ends at $\thvar$.  
Suppose there are $m \ge 0$ \nfail s between $\tvarp$ and $\thvarm$. 
Since $\bdata(\tvarp) < \bdata$ and $\bdata(\thvarm) = \bdata(\tvarp) - m$, 
\[ \bdata(\thvarp) = \bdata(\thvarm)+1 =  \bdata(\tvarp) +1 - m \le \bdata - m \]
and thus $\xvar_0$ has at least
\[ \ndata - m = \kdata +\bdata+ \rpdata -1 - m \ge 
\kdata + \bdata(\thvarp) + (\rpdata-1) \]
stored fragments at time $\thvarp$.  Furthermore,  $\xvar_j$ has at least 
\[ \kdata +\bdata(\thvarm)+ j = 
\kdata + \bdata(\thvarp) + (j-1)  \] 
stored fragments at time $\thvarp$
for $j = 1,\ldots, \rpdata-1$. The objects are logically reordered at time $\thvar$,
i.e., for $j = 1,\ldots, \rdata-1$, the old $\xvar_j$ becomes the new 
$\xvar_{j-1}$, and the old $\xvar_0$ becomes the new $\xvar_{\rpdata-1}$,
and the invariant is reestablished.

The duration of a repair step is set to
$\frac{1-\epsliq'}{\lambda \cdot \Ndata}.$ Thus,
\[ \rrate \le \frac{1-\beta}{(1-\epsliq) \cdot \beta} 
\cdot \lambda \cdot \Ndata\cdot \clen. \]

We now prove that with high probability the \srcdata\ is recoverable
over $\Mdata >> \Ndata$ \nfail s.  For the \srcdata\ to be unrecoverable, there must
be a \nfail\ at $\tvar$ where $\bdata(\tvarp) = \bdata-1$, 
another \nfail\ at $\thvar > \tvar$ where $\bdata(\thvarp) < 0$,
and repair steps are continually executing from $\tvar$ to $\thvar$. 

For $i \ge 1$, let $\Qvar_i$ be an independent exponential random variable 
with rate $\lambda \cdot \Ndata$.  Then, the probability the \srcdata\ is
unrecoverable within $\Mdata$ \nfail s is at most
\begin{equation}
\label{liq poisson eq}
\sum_{i=1}^{\Mdata} \sum_{m=\bdata}^{\Mdata}
\Prob{\sum_{j=1}^{m} \Qvar_j \le \frac{1-\epsliq'}{\lambda \cdot \Ndata}
\cdot (m - \bdata)},
\end{equation}
since $\frac{1-\epsliq'}{\lambda \cdot \Ndata}\cdot (m - \bdata)$ is the
time it takes to execute $m-\bdata$ repair steps, and the \srcdata\ is unrecoverable
only if the number of \nfail s exceeds the number of repair steps by at least $\bdata$
starting after a \nfail\ where repair steps initiate execution.

Let $\mu_m = \Exp{\sum_{j=1}^{m} \Qvar_j} = 
\frac{m}{\lambda \cdot \Ndata}$.
With 
\[ \theta = \epsliq' + \frac{\bdata \cdot (1-\epsliq')}{m}, \]
\[ \frac{1-\epsliq'}{\lambda \cdot \Ndata}
\cdot (m - \bdata) = (1-\theta) \cdot \mu_m. \]

From Theorem~5.1 of \cite{Janson17}, Equation~\eqref{liq poisson eq} is at most
\[ \Mdata \cdot \sum_{m=\bdata}^{\Mdata}
e^{-\frac{\lambda \cdot \Ndata \cdot \mu_m \cdot \theta^2}{2}} 
\le \Mdata \cdot \sum_{m=\bdata}^{\Mdata}
e^{-\frac{1}{2} \cdot 
\left(m \cdot \epsliq'^2 + \frac{\bdata^2\cdot (1-\epsliq')^2}{m}\right)}.  \]
It can be shown that for any $m$, 
\[ m \cdot \epsliq'^2 + \frac{\bdata^2 \cdot (1-\epsliq')^2}{m}
\ge 2 \cdot \epsliq' \cdot \bdata \cdot (1-\epsliq'). \]
Since $\bdata > \epsliq' \cdot \rdata$, Equation~\eqref{liq poisson eq} is at most
\[ \delliq = \Mdata^2 \cdot e^{-\epsliq'^2 \cdot (1-\epsliq')\cdot \rdata}. \]
\qed
\end{proof}

Although the \liqrepairer\ algorithm and analysis described in
\Liqpoissonthm\ is presented because it is self-contained and
relatively simple to understand, the algorithm and analysis 
in~\cite{Luby16} provides much tighter bounds.
Furthermore, \cite{Luby16} provides a much more sophisticated 
algorithm that dynamically and automatically adjusts the repair rate
as the \nfail\ rate fluctuates. 

\section{\Advliqrepairer}
\label{advliqrepairer sec}

For small $\beta$, the \liqrepairer\ read rate Equation~\eqref{liq upper eq}
in \Liqpoissonthm\ 
is around a factor of two larger than 
the \advliqrepairer\ read rate Equation~\eqref{aliq upper eq} in \ALiqpoissonthm\ 
described below in Section~\ref{aliq poisson sec}.
The intuition for this gap in optimality for the 
\liqrepairer\ is that although the maximum number of stored
fragments per object is $\ndata = \kdata + \rdata$, 
the average number of stored fragments per object is around 
$\kdata + \rdata/2$, i.e., only $\kdata+j+1$ 
stored fragments 
are needed for $\xvar_j$, for $j = 0,\ldots,\rdata-1$.
Thus $\frac{\beta}{2}$ of the allocated $\beta$ 
\storeoverhead\ is not in use at each time,
but the amount of data read by the \liqrepairer\ 
per \nfail\ is inversely proportional to $\beta$. 

The redundancy per object for the \advliqrepairer\ 
varies similar to that for the \liqrepairer.
The intuition behind the better bound for the 
\advliqrepairer\ is that fragments are generated and
stored using a strategy that allows the unequal object redundancy 
to be distributed equally among the nodes at each time,
while maintaining the property that the amount of data read by the
repairer per \nfail\ is inversely proportional to the maximum redundancy per object.
The \advliqrepairer\ uses a maximum of around 
$2\cdot \beta$ redundancy per object, but the average redundancy per object is $\beta$,
and this redundancy is equally distributed among the nodes at each time. 
Thus, the amount of data read per \nfail\ by the repairer 
is inversely proportional to $2 \cdot \beta$ in Equation~\eqref{aliq upper eq} 
of \ALiqpoissonthm\ when the \storeoverhead\ is $\beta$.

\subsection{\Advliqrepairer\ with periodic failures}
\label{aliq periodic sec}

As the core for proving the main algorithmic results, we introduce the
\advliqrepairer\ and prove upper bounds with respect to any \nfseq\ 
with a periodic \timeseq. The \advliqrepairer\ is based on~\cite{Luby16}.

\vspace{.1in}
\begin{ALiqperiodictheorem}
\label{ALiqperiodic theorem}
Fix $\beta > 0$ and let $(1-\beta) \cdot \Ndata \cdot \clen$
be the size of the \srcdata.
An \advliqrepairer\ with respect to any \nfseq\ with 
a periodic timing sequence
ensures the \srcdata\ is always recoverable.
The repairer read data per \nfail\ is at most
$\rlen \approx  \frac{1+2 \cdot \beta}
{2 \cdot \beta} \cdot \clen$,
the repairer write data per \nfail\ is at most
$\wlen \approx (2-\beta) \cdot \clen$.
\end{ALiqperiodictheorem}

\begin{proof}
The \advliqrepairer\ description follows.
Let $\kdata = \Ndata-1$ be the number 
of source fragments in an object.
Let $\rdata > 0$, and let $\ndata = \Ndata + \rdata$
be the total number of possible fragments for an object.
The \srcdata\ $\xvar$ is partitioned into $\Ndata$
groups of objects
\[
\xseq_0,\ldots,\xseq_{\Ndata-1},
\]
where, for $i=0,\ldots,\Ndata-1$, group $\xseq_i$ 
consists of an ordered set of $\rdata$ equal-size objects, i.e.,
\[ \xseq_i = \xvar_{i,0},\ldots,\xvar_{i,\rdata-1}.
\]

There are $\Ndata$ \primary\ EFIs  
\[ \fvar_0,\ldots,\fvar_{\Ndata-1}, \]
each assigned to one of the $\Ndata$ nodes,
where initially $\fvar_i = i$ for all $i = 0,\ldots,\Ndata-1$.
The storer uses an erasure code to generate and store
\primary\ fragment $\fvar_\ell=\ell$ 
for $\xvar_{i,j}$ at node $\ell$,
for $\ell = 0,\ldots,\Ndata-1$, $i= 0,\ldots, \Ndata-1$, 
$j= 0,\ldots, \rdata-1$.
Generally, EFI $\fvar_i$ is assigned 
to node $i$ and the \primary\ fragment
$\fvar_i$ for each object is stored at node $i$.

There is an ordered set of $\rdata$ \helper\ EFIs
\[ \hvar_{0},\ldots,\hvar_{\rdata-1}. \]
Initially, $\hvar_i = i + \Ndata$ 
for all $i = 0,\ldots,\rdata-1$.
\Helper\ fragments for the group $\xseq_i$ are stored at node $i$.
The storer generates and stores $j+1$ \helper\ fragments 
$\hvar_0=\Ndata,\ldots,\hvar_{j}=\Ndata+j$ 
for $\xvar_{i,j}$ at node $i$, 
for  $i= 0,\ldots, \Ndata-1$, 
$j = 0,\ldots, \rdata-1$.
Generally at time $\tvar$, 
the $j+1$ \helper\ fragments 
$\hvar_0,\ldots,\hvar_{j}$ 
for $\xvar_{i,j}$ are stored at node $i$, 
for  $i= 0,\ldots, \Ndata-1$, $j = 0,\ldots, \rdata-1$.

Each node stores $\Ndata \cdot \rdata$ \primary\ fragments and 
$\frac{\rdata \cdot (\rdata+1)}{2}$
\helper\ fragments, and thus the fragment size is
\begin{equation}
\label{flen eq}
\flen = \frac{\clen}{\rdata \cdot \left(\Ndata + \frac{\rdata+1}{2}\right)},
\end{equation}
 and the total number of stored fragments for all $\Ndata$ nodes is 
\[ \Ndata \cdot \rdata \cdot  \left(\Ndata + \frac{\rdata+1}{2} \right). \]
The total number of \srcdata\ fragments is
\[ \Ndata \cdot \rdata \cdot \kdata = \Ndata \cdot \rdata \cdot (\Ndata-1), \]
and thus the \storeoverhead\ is
\[ \beta = \frac{\rdata+3}{2 \cdot \Ndata + \rdata + 1}.\]
Note that for large $\Ndata$, 
\[\rdata \approx \frac{2 \cdot \beta \cdot \Ndata}{1-\beta}.\]

The following invariants are established by the storer and maintained
by the repairer for each node $i=0,\ldots,\Ndata-1$ at time $\tvarm$, 
where $\tvar$ is the time of a \nfail. 
\begin{itemize}

\item
\Primary\ fragment $\fvar_{i}$ for $\xvar_{\ell,j}$ is stored at node $i$
at time $\tvarm$, for $\ell = 0,\ldots,\Ndata-1$,  $j = 0,\ldots, \rdata-1$.

\item
\Helper\ fragments $\hvar_0,\ldots,\hvar_j$
for $\xvar_{i,j}$ are stored at node $i$ at time $\tvarm$,
for $j = 0,\ldots, \rdata-1$.  

\end{itemize}

From the invariants, all $\Ndata \cdot \rdata$ 
objects can be recovered from the \primary\ 
fragments stored at the nodes at time $\tvarp$, 
and thus the \srcdata\ always recoverable.

Suppose node $i$ fails at time $\tvar$ 
and is replaced by a node with all \bitval s set to zeroes.
To reestablish the invariants,
the repairer executes the following repair step for node $i$ 
starting at time $\tvarp$ and ending at a time $\thvar$ 
prior to the next \nfail.  

At time $\tvarp$ when the repair step begins, 
let $\hvar_0,\ldots,\hvar_{\rdata-1}$
be the ordered set of \helper\ EFIs, 
let $\fvar_0,\ldots,\fvar_{\Ndata-1}$
be the \primary\ EFIs,
and, for $\ell=0,\ldots,\Ndata-1$, let $\xvar_{\ell,0},\ldots,\xvar_{\ell,\rdata-1}$
be the ordering of $\xseq_\ell$.

Similarly, at time $\thvar$ when the repair step ends, 
let $\hhvar_0,\ldots,\hhvar_{\rdata-1}$
be the ordered set of \helper\ EFIs, 
let $\fhvar_0,\ldots,\fhvar_{\Ndata-1}$
be the \primary\ EFIs,
and, for $\ell=0,\ldots,\Ndata-1$, let $\xhvar_{\ell,0},\ldots,\xhvar_{\ell,\rdata-1}$
be the ordering of $\xseq_\ell$.

\begin{algorithm}
\caption{\em Repair step: for $i$}
\label{repair step alg}
\begin{algorithmic}
\State{\em Generate \helper s: for $i$}
\State{Set $\hhvar_0,\ldots,\hhvar_{\rdata-1} \leftarrow \hvar_1,\ldots,\hvar_{\rdata-1}, \fvar_i$}
\State{Set $\fhvar_i = \hvar_0$}
\For {$\ell = 0,\ldots,\Ndata-1$}
\State{\em Move \helper s: from $\ell$ to $i$} 
\State{\em Update \helper s: for $\ell$} 
\EndFor
\end{algorithmic}
\end{algorithm}

\begin{algorithm}
\caption{\em Generate \helper s: for $i$}
\label{generate helper alg}
\begin{algorithmic}
\For{$j=0,\ldots,\rdata-1$}
\State {Generate \helper\ fragments $\hvar_0,\ldots,\hvar_{j}$
for $\xvar_{i,j}$} \EndFor
\end{algorithmic}
\end{algorithm}

\begin{algorithm}
\caption{\em Move \helper s: from $\ell$ to $i$}
\label{move alg}
\begin{algorithmic}
\For{$j = 0,\ldots,\rdata-1$}
\State{Move fragment $\hvar_0 = \fhvar_i$ for $\xvar_{\ell,j}$ from node $\ell$ to node $i$}\EndFor
\end{algorithmic}
\end{algorithm}

\begin{algorithm}
\caption{\em Update \helper s: for $\ell$}
\label{update alg}
\begin{algorithmic}
\State{Set $\xhvar_{\ell,0},\ldots,\xhvar_{\ell,\rdata-1} 
\leftarrow \xvar_{\ell,1},\ldots,\xvar_{\ell,\rdata-1}, \xvar_{\ell,0}$}
\For{$j = 0,\ldots,\rdata-1$}
\State{Generate \helper\ fragments $\hhvar_0,\ldots,\hhvar_{\rdata-1}$
for $\xhvar_{\ell,\rdata-1}$}\EndFor
\end{algorithmic}
\end{algorithm}

Thus, by time $\thvar$ when the repair step ends, the first \helper\ EFI $\hvar_0$ 
becomes the \primary\ EFI for the replacement node $i$, the
remaining \helper\ EFIs $\hvar_1,\ldots,\hvar_{\rdata-1}$ 
are shifted forward by one position, and the \primary\ EFI $\fvar_i$
becomes the last \helper\ EFI.  
Thus, 
\[ \hhvar_0,\ldots,\hhvar_{\rdata-1} = \hvar_1,\ldots,\hvar_{\rdata-1}, \fvar_i, \]
$\fhvar_i = \hvar_0$, and $\fhvar_\ell = \fvar_\ell$ for $\ell \not= i$.
Furthermore, for  $\ell=0,\ldots,\Ndata-1$, all of the objects in $\xseq_\ell$ 
have cyclic shifted by one position forward, i.e.,
\[ \xhvar_{\ell,0},\ldots,\xhvar_{\ell,\rdata-1} = 
\xvar_{\ell,1},\ldots,\xvar_{\ell,\rdata-1}, \xvar_{\ell,0}. \]

It can be verified the invariants are reestablished at $\thvar$
when the repair step completes.
The number of fragment reads for {\em Generate \helper s}
is $(\Ndata-1) \cdot \rdata$ and the number of fragment writes is 
$\frac{\rdata \cdot (\rdata+1)}{2}$.
The number of fragment reads for 
{\em Move \helper s} is $\rdata$ 
and the number of fragment writes is $\rdata$.
The number of fragment reads for 
{\em Update \helper s} is $\Ndata -1$ 
and the number of fragment writes is $\rdata$.
Thus, using Equation~\eqref{flen eq}, the amount of read data for
the execution of {\em Repair step} is
\begin{equation*}
\rlen \le \frac{\Ndata \cdot (\Ndata + 2 \cdot \rdata ) }
{\rdata \cdot \left(\Ndata + \frac{\rdata+1}{2}\right)}\cdot \clen,
\end{equation*}
and the amount of written data is
\begin{equation*}
\wlen = \frac{2 \cdot \Ndata+ \frac{\rdata+1}{2}}
{\Ndata + \frac{\rdata+1}{2}} \cdot \clen.
\end{equation*}

For large $\Ndata$, 
\[\rlen\approx \frac{(1 + 3 \beta) \cdot (1-\beta)}{2 \cdot \beta} \cdot \clen
\le \frac{1+2 \cdot \beta}{2 \cdot \beta} \cdot \clen \]
and 
\[ \wlen \approx (2-\beta) \cdot \clen.\]
\qed
\end{proof}

\subsection{\Advliqrepairer\ with Poisson failures}
\label{aliq poisson sec}

The \advliqrepairer\ of \ALiqpoissonthm\ is based on~\cite{Luby16}.

\vspace{.1in}
\begin{ALiqpoissontheorem}
\label{ALiqpoisson theorem}
Fix $\beta > 0$ and let $(1-\beta) \cdot \Ndata \cdot \clen$
be the size of the \srcdata.
For  $\epsliq > 0$ let
\begin{equation*}
\delliq = e^{-\frac{\epsliq^2 \cdot \left(1-\frac{\epsliq}{2}\right)
\cdot \beta \cdot \Ndata}{4 \cdot (2 \cdot \beta + 1)}}.
\end{equation*}
Let the \nfseq\ be a \Ptdist\ with rate $\lambda$.
With probability at least $1-\Mdata^2\cdot \delliq$, an
\advliqrepairer\ can maintain recoverability of 
the \srcdata\ for $\Mdata$ \nfail s 
when the peak repairer read rate
\begin{equation}
\label{aliq upper eq}
\rrate = \frac{1-\beta}{1-\epsliq'}
\cdot \left(1+\frac{1}{2 \cdot (\beta-\epsliq')}\right) 
\cdot \lambda \cdot \Ndata\cdot \clen.  
\end{equation}
\end{ALiqpoissontheorem}

\begin{proof}
The \advliqrepairer\ is essentially the same as that described in
the proof of \ALiqperiodicthm, with the differences described below.
Let $\epsliq' = \epsliq/2$. 
Set $\bdata = \epsliq' \cdot \Ndata+1$,
\[ \kdata = \Ndata - \bdata = \Ndata \cdot (1-\epsliq') - 1.\]
Thus, the \storeoverhead\ is
\[ \beta = 1 - \frac{\Ndata - \bdata}{\Ndata + \frac{\rdata+1}{2}}
= \frac{\rdata + 1 + 2 \cdot \bdata}{2 \cdot \Ndata + \rdata + 1}. \]
It can be verified that 
\[ \rdata \le \frac{2 \cdot \Ndata \cdot (\beta - \epsliq')}{1-\beta}. \]

Let $\bdata(\tvar)$ be a counter that is updated
as follows, where initially $\bdata(0) = \bdata$. 
\begin{itemize}
\item 
$\bdata(\tvarp) = \bdata(\tvarm)-1$ if there is a \nfail\ at $\tvar$. 
\item
$\bdata(\tvarp) = \min\{\bdata(\tvarm)+1,\bdata\}$ 
if a repair step ends at $\tvar$.
\end{itemize}

A system invariant is that at time $\tvar$ there is a node set
$\Nsetp(\tvar) \subseteq \{0,\ldots,\Ndata-1\}$
such that $\abs{\Nsetp(\tvar)} \ge \kdata + \bdata(\tvar)$
and the following invariants hold for each node $i \in \Nsetp(\tvar)$:
\begin{itemize}

\item
\Primary\ fragment $\fvar_{i}$ for $\xvar_{\ell,j}$ is stored at node $i$
at time $\tvar$, for $\ell = 0,\ldots,\Ndata-1$,  $j = 0,\ldots, \rdata-1$.

\item
\Helper\ fragments $\hvar_0,\ldots,\hvar_j$
for $\xvar_{i,j}$ are stored at node $i$ at time $\tvar$,
for $j = 0,\ldots, \rdata-1$.  

\end{itemize}
These invariants are established by the storer and maintained by the repairer.
From these invariants, the \srcdata\ is recoverable up to time $\tvar$
as long as $\bdata(\thvar) \ge 0$ for all $\thvar \le \tvar$.

The invariants are reestablished at time $\tvarp$ after a \nfail\ at time $\tvar$, 
since $\abs{\Nsetp(\tvarp)} \ge \abs{\Nsetp(\tvarm)}-1$.

A repair step is similar to the repair step in the proof of \ALiqperiodicthm,
with some important differences.
No repair step is executing at time $\tvar$ if $\bdata(\tvar) = \bdata$.  
If there is a \nfail\ at time $\thvar$ when $\bdata(\thvarm) = \bdata$, 
then $\bdata(\thvarp) = \bdata-1$ and a repair step is initiated.
Repair steps are executed sequentially, i.e., at most one repair step
is executing at any time.  As soon as a repair step completes,
a next repair step can initiate.

Suppose the following {\em Repair step} starts 
at time $\tvar$ and ends at time $\thvar$, where $i \notin \Nsetp(\tvar)$.
{\em Generate \helper s},
{\em Move \helper s}, and
{\em Update \helper s} are the same as described
in the proof of \ALiqperiodicthm.

\begin{algorithm}
\caption{\em Repair step: for $i$}
\label{poisson repair step alg}
\begin{algorithmic}
\For {$\ell = 0,\ldots,\Ndata-1$}
\While{{\em Move \helper s: from $\ell$ to $i$} is incomplete}
\While{node $\ell$ does not have \helper s}
\State{\em Generate \helper s: for $\ell$}
\EndWhile
\While{node $\ell$ has \helper s}
\State{\em Move \helper s: from $\ell$ to $i$}
\EndWhile
\EndWhile
\EndFor
\State{Set $\hhvar_0,\ldots,\hhvar_{\rdata-1} \leftarrow \hvar_1,\ldots,\hvar_{\rdata-1}, \fvar_i$}
\State{Set $\fhvar_i = \hvar_0$}
\For {$\ell = 0,\ldots,\Ndata-1$}
\State{\em Update \helper s: for $\ell$} 
\EndFor
\end{algorithmic}
\end{algorithm}

The primary change to {\em Repair step} from the proof of
\ALiqperiodicthm\ is the following.
If node $\ell$ does not have \helper\ fragments
just before the execution of {\em Move \helper s}
then {\em Generate \helper s} is executed first to generate 
\helper\ fragments.
If node $\ell$ fails during the execution of either 
{\em Generate \helper s} or 
{\em Move \helper s}, then
{\em Generate \helper s} is executed again. This repeats until  
{\em Move \helper s} completes.
Thus, the amount of data read during the execution of {\em Repair step} can vary.

Suppose there are $m \ge 0$ \nfail s between $\tvarp$ and $\thvarm$.
Then $\bdata(\thvarp) = \bdata(\tvarp) - m + 1$.
If node $i$ is not among the failing nodes then $i \in \Nsetp(\thvarp)$,
and  since $i \not\in \Nsetp(\tvarp)$ then
$\abs{\Nsetp(\thvarp)} \ge \abs{\Nsetp(\tvarp)}-m+1$.
If node $i$  is among the failing nodes 
then $i \not\in \Nsetp(\thvarp)$, and
since $i \not\in \Nsetp(\tvarp)$ then
$\abs{\Nsetp(\thvarp)} \ge \abs{\Nsetp(\tvarp)}-(m-1)$.
It follows that the invariants are reestablished at $\thvarp$.

The amount of data read per {\em Repair step} for 
the $\Ndata$ executions of {\em Move \helper s} and
{\em Update \helper s} is at most
\begin{equation}
\label{move update eq}
(1-\beta) \cdot \left(1 + \frac{1}{2 \cdot (\beta-\epsliq')} \right)\cdot \clen.
\end{equation}

The amount of data read for one execution of {\em Generate \helper s}
within {\em Repair step} is at most
\begin{equation}
\label{generate eq}
(1-\beta) \cdot  \clen.
\end{equation}

Let
\[ \rrate = \frac{1-\beta}{1-\epsliq'}
\cdot \left(2+\frac{1}{2 \cdot (\beta-\epsliq')}\right) 
\cdot \lambda \cdot \Ndata\cdot \clen. \]

Then, using Equation~\eqref{move update eq} and Equation~\eqref{generate eq},
 the time to complete the $\Ndata$ executions of the 
 {\em Move \helper s} and
{\em Update \helper s} portions of a repair step, plus the time
for one execution of {\em Generate \helper s}, is at most
\begin{equation}
\label{time move update eq}
\frac{1-\epsliq'}{\lambda \cdot \Ndata},
\end{equation}
and, from Equation~\eqref{generate eq}, 
the time to complete one execution of {\em Generate \helper s}
is at most
\begin{equation}
\label{time generate eq}
\frac{1-\epsliq'}{\lambda \cdot \Ndata} 
\cdot \frac{2 \cdot \beta}{2 \cdot \beta + 1}.
\end{equation}

The time to complete $m - \bdata$ {\em Repair steps} is at most
\begin{equation}
\label{time repair eq}
\frac{1-\epsliq'}{\lambda \cdot \Ndata}
\cdot \left(m - \frac{\bdata}{2 \cdot \beta + 1}\right) ,
\end{equation}
since $\frac{1-\epsliq'}{\lambda \cdot \Ndata}\cdot (m - \bdata)$ 
is the time it takes to execute the {\em Move \helper s} and
{\em Update \helper s} portions of $m - \bdata$ {\em Repair steps}, 
plus the time for $m - \bdata$ executions of {\em Generate \helper s}, 
and there can be at most $m$ executions of {\em Generate \helper s} 
when there are $m$ \nfail s, and 
$\frac{1-\epsliq'}{\lambda \cdot \Ndata} \cdot
\frac{\bdata \cdot 2 \cdot \beta}{2 \cdot \beta + 1}$ 
is at most the time for $\bdata$ additional executions of {\em Generate \helper s}.

The remainder of the proof is similar to the last part of the proof of \Liqpoissonthm.
We prove that with high probability the \srcdata\ is recoverable
over $\Mdata >> \Ndata$ \nfail s.  For the \srcdata\ to be unrecoverable, there must
be a \nfail\ at $\tvar$ where $\bdata(\tvarp) = \bdata-1$, 
another \nfail\ at $\thvar > \tvar$ where $\bdata(\thvarp) < 0$,
and repair steps are continually executing from $\tvar$ to $\thvar$. 

For $i \ge 1$, let $\Qvar_i$ be an independent exponential random variable 
with rate $\lambda \cdot \Ndata$.  Then, from Equation~\eqref{time repair eq},
the probability the \srcdata\ is unrecoverable within $\Mdata$ \nfail s is at most
\begin{equation}
\label{aliq poisson eq}
\sum_{i=1}^{\Mdata} \sum_{m=\bdata}^{\Mdata}
\Prob{\sum_{j=1}^{m} \Qvar_j \le \frac{1-\epsliq'}{\lambda \cdot \Ndata}
\cdot \left(m - \frac{\bdata}{2 \cdot \beta + 1}\right)}.
\end{equation}

Let $\mu_m = \Exp{\sum_{j=1}^{m} \Qvar_j} = 
\frac{m}{\lambda \cdot \Ndata}$.
With 
\[ \theta = \epsliq' + \frac{\bdata \cdot (1-\epsliq')}{m \cdot (2 \cdot \beta + 1)}, \]
\[ \frac{1-\epsliq'}{\lambda \cdot \Ndata}
\cdot\left (m - \frac{\bdata}{2 \cdot \beta + 1}\right) = (1-\theta) \cdot \mu_m. \]

From Theorem~5.1 of \cite{Janson17}, Equation~\eqref{aliq poisson eq} is at most
\[ \Mdata \cdot \sum_{m=\bdata}^{\Mdata}
e^{-\frac{\lambda \cdot \Ndata \cdot \mu_m \cdot \theta^2}{2}} 
\le \Mdata \cdot \sum_{m=\bdata}^{\Mdata}
e^{-\frac{1}{2} \cdot 
\left(m \cdot \epsliq'^2 + \frac{\bdata^2\cdot 
(1-\epsliq')^2}{m \cdot (2 \cdot \beta+1)^2}\right)}.  \]
It can be shown that for any $m$, 
\[ m \cdot \epsliq'^2 + \frac{\bdata^2 \cdot (1-\epsliq')^2}{m\cdot (2 \cdot \beta+1)^2}
\ge \frac{2 \cdot \epsliq' \cdot \bdata \cdot (1-\epsliq')}{2 \cdot \beta + 1}. \]
Since $\bdata > \epsliq' \cdot \rdata$, Equation~\eqref{aliq poisson eq} is at most
\[ \delliq = \Mdata^2 \cdot e^{-\frac{\epsliq'^2 \cdot (1-\epsliq')\cdot \rdata}
{2 \cdot \beta + 1}}. \]
\qed
\end{proof}

As $\Ndata$ grows  and $\beta$ goes to zero, $\epsilon$ can go to zero,
and Equation~\eqref{aliq upper eq}
asymptotically approaches
\begin{equation}
\label{alglimit eq}
\rrate \le \frac{1}{2 \cdot \beta} \cdot \lambda \cdot \Ndata \cdot \clen.
\end{equation}

\section{Related work}
\label{related work sec}

The groundbreaking research of Dimakis et. al., described in~\cite{Dimakis07} 
and~\cite{Dimakis10}, is closest to our work:
An object-based distributed storage framework is introduced, 
and optimal tradeoffs between \storeoverhead\ and \lrepairer\ read rate are proved
with respect to repairing an individual object. 
We refer to the framework, the lower bounds, and the repairer
described in~\cite{Dimakis07} and~\cite{Dimakis10} as the {\em \Regframe}, 
the {\em \Reglower s}, and the {\em \Regrepairer}, respectively. 

The \Regframe\ models repair of a single lost fragment, 
and is applicable to \tradrepair\ of a single object. 
The \Regframe\ is based on $(n, k, d, \alpha, \gamma)$: 
$n$ is the number of fragments for the object (each stored at a different node); 
$k$ is the number of fragments from which the object must be recoverable; 
$d$ is the number of fragments used to generate a lost fragment at a new node 
when a node fails; $\alpha$ is the fragment size;
and $\gamma/d$ is the amount of data generated from 
each of $d$ fragments needed to generate a fragment at a new node.

\subsection{\Regrepairer s}

We consider Minimum Storage Regenerating (MSR) settings (the object size is $\alpha \cdot k$), 
and fix $d= n-1$, as these are typically thought of as the most practical settings that
minimize storage overhead and minimize repairer traffic.

The \Regrepairer~\cite{Dimakis07}, \cite{Dimakis10}
is a \lrepairer\ (Section~\ref{repairer sec}).
When a node fails, at each of the $n-1$ remaining nodes storing a fragment for the object
a network coding function is applied locally to the fragment at that node 
to generate $\frac{\gamma}{n-1}$ \bitval s that are read over the node interface
by the repairer, which uses all \bitval s read over interfaces from the nodes 
to generate and store a fragment at the new node.
The $\frac{\gamma}{n-1}$ \bitval s read over the node interface by the repairer 
from each of the $n-1$ nodes is counted as data read by the repairer,
whereas accessing the fragment of size $\alpha$ to locally generate the 
$\frac{\gamma}{n-1}$ \bitval s at each node is not counted as
data read by the repairer.
At the optimal setting that minimizes the amount of data read over node 
interfaces by the repairer in this model,
\[\alpha \approx (n-k) \cdot \frac{\gamma}{n-1},\] 
thus the amount of data locally accessed at each node is $n-k$ 
times the size of the data counted as read over the interface from the node.

The \Regrepairer~\cite{Kumar16} is more advanced.
When a node fails, at each of the $n-1$ remaining nodes
a selected subset of $\frac{\gamma}{n-1}$ \bitval s of the fragment 
is read over the interfacee from that node by the repairer,
which uses all received \bitval s over interfaces from the nodes 
to generate and store a fragment at the new node.
This construction can use Reed-Solomon codes,
and is efficient for small values of $k$ and $n$.

Each fragment is partitioned into sub-fragments for the \Regrepairer~\cite{Kumar16},
and the number of sub-fragments provably grows quickly as $n-k$ grows or 
as the \storeoverhead\ $\beta=\frac{n-k}{n}$ approaches zero.  
For example there are $1024$ sub-fragments per fragment for $k=16$ and $n=20$,
and the \Regrepairer\ generates a fragment at a new node
from receiving $256$ non-consecutive sub-fragments over an interface 
from each of $19$ nodes.
Accessing many non-consecutive sub-fragments directly from a node is sometimes efficient,
in which case the amount of data read over interfaces from nodes
is equal to the amount of accessed data used to create the read data, 
but typically it is more efficient to access an entire fragment at a node 
and locally select the appropriate sub-fragments to send 
over the interface from the node to the repairer, 
in which case the amount of data locally accessed at each node is $n-k$ 
times the size of the data counted as read over the interface from the node.

In contrast to the \Regrepairer s~\cite{Dimakis07}, \cite{Dimakis10}, \cite{Kumar16},
the repairer described in the proof of \ALiqpoissonthm\ 
directly reads unmodified data over interfaces from nodes, 
e.g. using HTTP, thus the upper bound on the amount of data 
read over interfaces from nodes also accounts for all data accessed from nodes.

When \Regrepairer s~\cite{Dimakis07}, \cite{Dimakis10}, \cite{Kumar16}
are used to repair all objects in a system, the ratio of the amount of data read
per node failure to the capacity of a node is approximately $\frac{1}{\beta}$
with respect to any \nfseq\ with a periodic timing sequence,
which is around a factor of two above the lower bound of approximately
$\frac{1}{2 \cdot \beta}$ from Equation~\eqref{asymp read rate eq}.

When \Regrepairer s~\cite{Dimakis07}, \cite{Dimakis10}, \cite{Kumar16} 
with small $n$ and $k$ are used to repair objects as $\Ndata$ grows,
the needed peak repairer read rate with respect to a \Ptdist\ 
grows far beyond the lower bound average \rrepairrate\ of \Poissonthm. 
This is because there is a good chance that multiple \nfail s occur over a small interval
of time, implying that repair for \nfail s must occur in a very short
interval of time.
In contrast, for the repairer described in the proof of \ALiqpoissonthm,
the peak repairer read rate essentially matches the lower bound
average \rrepairrate.

Thus, \Regrepairer s ~\cite{Dimakis07}, \cite{Dimakis10}, \cite{Kumar16}
minimize the peak repairer read rate with respect to the \Regframe\
for repairing individual objects using \tradrepair, but
they do not provide optimal peak repairer read rate at the system level
when used to store and repair objects.

\subsection{\Reglower s}

\Reglower s on the \lrepairer\ read data rate prove necessary
 conditions on the \Regframe\ parameters 
to ensure than an individual object remains recoverable when using \tradrepair.
The bounds are based on a specially constructed acyclic graph, which corresponds to a specially constructed \nfseq,
and does not show for example a lower bound for uniformly chosen \nfail s.
Also, the lower bounds are not extendable to non-trivial \timeseq s, e.g., \Ptdist s.

Consider applying the \Regframe\ at the system level across all objects, 
e.g.,  $n = \Ndata$, and, for MSR settings, $k=\Kdata$ and 
$\alpha = \frac{\xlen}{k}$ for \srcdata\ of size $\xlen$.
The following two examples show this system level \Regframe\ 
does not capture the properties of existing repairer algorithms,
thus the \Reglower s do not provide system level lower bounds.

A system level \Regframe\ requirement would be 
that all \srcdata\ is recoverable from any $\Kdata$ of the $\Ndata$ nodes. 
However, as described in Section~\ref{practical sec}, \tradsystem s partition \srcdata\ 
into objects, and fragments for objects are distributed equally to all $\Ndata$ nodes,
thus \tradsystem s read data from almost all $\Ndata$ nodes to recover all \srcdata,
which violates this requirement. 
 
Another system level \Regframe\ requirement would be 
that data is only written to a node when it is added: 
Writing data incrementally to a node over time as nodes fail is not expressible.
\Liqsystem s write data incrementally to a node over a large number of \nfail s after the node
is added, which violates this requirement. 

\section{Future work}

There are many ways to extend this research, accounting for practical issues
in storage system deployments.

Failures in deployed systems can happen at a variable rate that is not known a priori.
For example, a new batch of nodes introduced into a deployment 
may have failure rates that are dramatically different than previous batches.
The paper~\cite{Luby16} introduces repair algorithms that automatically adjust to
fluctuating failure rates.

Both time and spatial failure correlation is common in deployed systems.
Failures in different parts of the system are not completely independent, 
e.g., racks of nodes fail concurrently, entire data centers go offline, 
power and cooling units fail, node outages occur due to rolling 
system maintenance and software updates, etc.
All of these events introduce complicated correlations between
failures of the different components of the system. 

Intermittent \nfail s are common in deployed 
systems, accounting for a vast majority (e.g., 90\%) of \nfail s.  
In the case of an intermittent node failure, the data stored at the node 
is lost for the duration of the failure, but after some period of time
the data stored on the node is available again once the node recovers 
(the period of time can be variable, e.g., ranging from a few seconds to days).
Intermittent failures can also affect entire data centers, a rack of nodes, etc. 

Repairing fragments temporarily unavailable due to 
transient \nfail s wastes network resources.  
Thus, a timer is typically set to trigger a fixed amount of time
after a node fails (e.g.,15 minutes), 
and the node is declared permanently failed and scheduled for repair if it has not recovered within the trigger time. 
Setting the trigger time can be tricky for a \tradsystem; a short trigger time can lead
to unnecessary repair, whereas a long trigger time can reduce reliability.
The paper~\cite{Luby16} provides simulations that highlight the
impact of setting the trigger time for different systems.

Data can silently be corrupted or lost without any notification to the repairer; 
the only mechanism by which a repairer may become aware of such corruption or 
loss of data is by attempting to read the data, i.e., data scrubbing.  
(The data is typically stored with strong checksums, so that the corruption 
or loss of data becomes evident to the repairer
when an attempt to read the data is made.)
For example, the talk~\cite{Cowling16} reports that 
read traffic due to scrubbing can be greater than all other read data traffic combined.
The paper~\cite{Luby16} provides simulations that highlight the
impact of silent corruption on different systems.

There can be a delay between when a node 
permanently fails and when a replacement node is
added.  For example, in many cases adding nodes is performed by robots,
or by manual intervention, and nodes are added in batches instead of individually.

It is important in many systems to distribute the repair 
evenly throughout the nodes and the network,
instead of having a centralized repairer.  
This is important to avoid CPU and network hotspots.
It can be verified that the algorithms 
described in Sections~\ref{liqrepairer sec} and~\ref{advliqrepairer sec}
can be modified to distribute the repair traffic smoothly among all nodes of the system.  
Based on this, it can be seen that distributed versions of the 
lower bounds and upper bounds asymptotically converge as the
\storeoverhead\ approaches zero.

Network topology is an important consideration in deployments, for example when objects are geo-distributed
to multiple data centers.  In these deployments, the available network bandwidth between different nodes
may vary dramatically, e.g., there may be abundant bandwidth available between nodes within the same data center,
but limited bandwidth available between nodes in different data centers.
The paper~\cite{Gopalan16} addresses these issues, 
and the papers~\cite{Gopalan12}, \cite{Huang12}
introduce some erasure codes that may be used in solutions to these issues.
An example of such a deployment is described in~\cite{Facebook14}.

Enhancing the distributed storage model by incorporating the elements described above
into the model and providing an analysis can be of value in understanding fundamental tradeoffs for practical systems.

\section{Conclusions}
We introduce a mathematical model of distributed 
storage that captures some of the relevant features of
practical systems, and prove tight bounds
on distributed storage \srcdata\ capacity.
Our hope is that the model and bounds will 
be helpful in understanding and 
designing practical distributed storage systems.

\section*{Acknowledgment}

I thank Roberto Padovani for consistently championing this research.  
I thank members of the Qualcomm systems team 
(Roberto, Tom Richardson, Lorenz Minder, Pooja Aggarwal)
for being great collaborators and for providing invaluable feedback 
on this work as it evolved.
  
I thank the Simons Institute at UC Berkeley for sponsoring 
the Information Theory program January-April 2015, 
as the participants in this program provided a lot of detailed information about work in 
distributed storage that helped understand the context of our research in general.
  
I thank Tom Richardson for taking the time to understand and provide 
improvements to this research, ranging from high level presentation suggestions
to simplifications of proofs.

I thank Tom and the organizers of the Shannon lecture series at UCSD for conspiring 
to invite me to give the Shannon lecture December 1, 2015 -- preparing the presentation for this lecture inspired 
thinking about a mathematical model analogous to Shannon's communication theory model.

I thank colleagues at Dropbox (in particular James Cowling), 
Microsoft Azure (in particular Cheng Huang and Parikshit Gopalan), 
Google (in particular Lorenzo Vicisano) and Facebook for sharing valuable
insights into operational aspects and potential issues with large scale deployed distributed storage systems.  


%
\appendices




%

\begin{IEEEbiographynophoto}{Michael G. Luby} is VP Technology, Qualcomm, Inc.
Research and development projects include Liquid distributed storage, 
LTE broadcast multimedia delivery, and DASH internet streaming.  
Mike earned a BSc in Applied Math from MIT and a PhD in Theoretical Computer Science from UC Berkeley. 
He founded Digital Fountain Inc. in 1999, and served as CTO until acquired by Qualcomm Inc. in 2009. 
Awards for his research include 
the IEEE Richard W. Hamming Medal, 
the ACM Paris Kanellakis Theory and Practice Award, 
the ACM Edsger W. Dijkstra Prize in Distributed Computing, 
the ACM SIGCOMM Test of Time Award, 
the IEEE Eric E. Sumner Communications Theory Award, 
the ACM SIAM Outstanding Paper Prize, 
the UC Berkeley Distinguished Alumni in Computer Science Award, 
and the IEEE Information Theory Society Information Theory Paper Award.  
He is a member of the National Academy of Engineering, and is an IEEE Fellow and an ACM Fellow.\end{IEEEbiographynophoto}








\end{document}